%% file: LATIN_Dynamics_SEBA_V2.tex
\RequirePackage{fix-cm}
\documentclass[smallextended]{svjour3}       
\smartqed  
\usepackage{graphicx}
\usepackage{tcolorbox}
\usepackage{graphicx}
\usepackage{empheq}
\usepackage{framed}
\usepackage{afterpage}
\usepackage{appendix}

\usepackage{tikz}
\usepackage{import}
\usepackage{float}
\usepackage{xcolor}
\usepackage{amsmath}
\usepackage{multirow}
\usepackage{amsmath,amsfonts,amssymb,amsthm}
\usepackage{mathtools}
\usepackage[linesnumbered,ruled]{algorithm2e}
\usepackage{mathtools}
\usepackage{mathrsfs}
\usepackage[super]{nth}
\usepackage{caption}
\usepackage{subcaption}
\usepackage{graphicx}
\usepackage{wrapfig}
\newcommand*\rfrac[2]{\mbox{\raisebox{1.5pt}{$\scriptscriptstyle{#1}$}}\hspace{-1pt}/\hspace{-1pt}\mbox{\raisebox{-1pt}{$\scriptscriptstyle{#2}$}}}

\usepackage{verbatim} 
\usepackage{longtable}

\usepackage{array}
\usepackage{supertabular}
\usepackage{longtable}
\usepackage{tabularx}
\usepackage{multirow}
\usepackage{colortbl}
\usepackage{fancyhdr}
\usepackage{xcolor}
\usepackage{pgf,tikz}
\usepackage{pgfplots, pgfplotstable}
\usetikzlibrary{fit,calc,positioning,decorations.pathreplacing,matrix}
\usetikzlibrary{arrows}
\usetikzlibrary{shapes}
\usetikzlibrary{chains}

\tikzstyle{load}   = [ultra thick,-latex]
\tikzstyle{stress} = [-latex]
\tikzstyle{dim}    = [latex-latex]
\tikzstyle{axis}   = [-latex,black]

\tikzstyle{isometric}=[x={(0.710cm,-0.410cm)},y={(0cm,0.820cm)},z={(-0.710cm,-0.410cm)}]
\tikzstyle{dimetric} =[x={(0.935cm,-0.118cm)},y={(0cm,0.943cm)},z={(-0.354cm,-0.312cm)}]
\tikzstyle{dimetric2}=[x={(0.935cm,-0.118cm)},z={(0cm,0.943cm)},y={(+0.354cm,+0.312cm)}]
\tikzstyle{trimetric}=[x={(0.926cm,-0.207cm)},y={(0cm,0.837cm)},z={(-0.378cm,-0.507cm)}]

\usepackage{multido}
\usepackage{calc}
\usepackage{fp}
\usepackage{xifthen}
\usepackage{xargs}
\usepackage{etoolbox}
\DeclareMathAlphabet{\mathpzc}{OT1}{pzc}{m}{it}
\usepackage{hyperref}
\usepackage{breakcites}  
\usepackage[utf8]{inputenc}
\usepackage{helvet}
\usepackage{geometry}
\usepackage{amssymb}
\usepackage{enumitem}
\usepackage{tabularx}
\usepackage{xcolor}
\usepackage[absolute,overlay]{textpos}
\usepackage{graphicx}
\usepackage{lipsum}
\usepackage{caption}
\usepackage{multicol}
\usepackage{hyperref}
\usepackage{afterpage}
\usepackage{setspace}
\usepackage{pgffor}
\usepackage{sistyle}
\usepackage{xfrac}

\newcommand{\CharboModif}[2]{\color{#1}{#2}\color{black}}

\import{./}{Colors}

\import{./}{Commandes} 		

\import{./}{Notations} 		

\begin{document}

\title{

The LATIN-PGD methodology to nonlinear dynamics and quasi-brittle materials for future earthquake engineering applications
}


\author{Sebastian Rodriguez \and Pierre-Etienne Charbonnel \and Pierre Ladevèze \and David Néron}



\institute{Sebastian Rodriguez \at
              \email{sebastian.rodriguez\_iturra@ens-paris-saclay.fr}           
 \and
Pierre-Etienne Charbonnel \at
\email{pierreetienne.charbonnel@cea.fr}
\and
Pierre Ladevèze \at
  \email{ladeveze@ens-paris-saclay.fr}
\and
David Néron \at
   \email{david.neron@ens-paris-saclay.fr}
}

\date{Received: date / Accepted: date}

\maketitle

\begin{abstract}

This paper presents a first implementation of the LArge Time INcrement (LATIN) method along with the model reduction technique called Proper Generalized Decomposition (PGD) for solving nonlinear low-frequency dynamics problems when dealing with a quasi-brittle isotropic damage constitutive relations. 
The present paper uses the Time-Discontinuous Galerkin Method (TDGM) for computing the temporal contributions of the space-time separate-variables solution of the LATIN-PGD approach, which offers several advantages when considering a high number of DOFs in time. The efficiency of the method is tested for the case of a 3D bending beam, where results and benchmarks comparing LATIN-PGD to classical time-incremental Newmark/Quasi-Newton nonlinear solver are presented. This work represents a first step towards taking into account uncertainties and carrying out more complex parametric studies imposed by seismic risk assessment.


\end{abstract}

\section{Introduction}\label{intro}


The design of civil engineering structures with respect to seismic risk is a very numerically demanding process and this trend is further accentuated when considering different facilities for which additional safety studies are imposed on the most sensitive components.
Indeed, the complexity and richness of the numerical models used to predict the often nonlinear behavior of structures generate computation times of several days for the simulation of a single seismic event using classical Newmark-like incremental methods.
Furthermore, assessing the margins and taking into account the variability of the parameters of the reference problem lead to make this numerical effort, no longer for the simulation of a single model, but of a family of models.
This work is thus dedicated to the derivation of a numerical strategy for computing reliable solutions for nonlinear dynamics in the low-frequency range (typical seismic inputs have a frequency content below $50Hz$) trying to minimize the associated computational cost.
In this paper, focus is made on concrete considering the predominant role it plays for earthquake-resistant constructions and civil engineering, but the proposed methodology is general and could be applied to any kind of material with nonlinear behavior.


Among the different strategies dedicated to solving parametric problems, some methods, released in the 2000's and currently booming, propose to use an ingredient referred to as \emph{model order reduction} which confers them a formidable numerical efficiency.
The main idea is to exploit the redundancy of information contained in the solution to propose a numerically efficient solution of the problem, which guarantees that the calculated approximation, called \emph{low-rank approximation}, stays close enough to the solution.
The solution of the reference problem is thus approached by a sum of $\PGDm$ terms where each of the terms is a product of functions with separate variables. The integer $\PGDm$ is called the rank of the approximation and in practice, the approximation space is constructed incrementally. Among those methods, one can cite the \emph{Certified Reduced Basis Method} (CRBM) and the \emph{Proper Generalized Decomposition} (PGD).
The CRBM methodology has been introduced initially in \cite{maday2002priori,prud2002reliable} for the resolution of elliptic parametric problems. 
The methodology resides in the sequential construction of a reduced basis of the solution space which enables, after projection of the governing equations on the latter, the accelerated ``online'' resolution of the parametric problem. 
The reduced basis itself is constructed ``offline'' after \emph{Proper Orthogonal Decomposition} (POD) of finite-element solutions associated to a sequence of parameters chosen in an optimal manner.
The underlying hypothesis for the online computation of the parametric solution is the affine dependence of the operators of the reference problem with respect to the parameter  vector; i.e. the bilinear forms involved must be written as a sum of products of forms with separated variables. The CRBM has then been adapted for parabolic problems and extended in the nonlinear range for the resolution of problems having any kind of parametric dependence, employing the \emph{Empirical Interpolation Method} (EIM)  for approximating operators with an affine sequence  (see \cite{grepl2005reduced,prud2002mathematical,quarteroni2011certified} and \cite{veys2014framework} for details).
Alternatively, the PGD was introduced in \cite{ladeveze1985famille} under the vocable ``radial approximation'' as one of the main ingredients of the LATIN method.
The LATIN method \cite{ladeveze1999nonlinear} proposes a general strategy for the resolution of nonlinear problems in mechanics involving an alternative sequence of \emph{nonlinear} and \emph{linear} steps. 
At each linear step, a global space-time problem expressing the equilibrium of the system must be solved; the PGD is used to provide a reliable and numerically economical low-rank approximation of the solution of this linear problem. Since its introduction in \cite{ladeveze1985famille}, the PGD has been the subject of numerous other publications for solving different types of linear problems with parametric dependence, following deterministic \cite{ammar2006new,ryckelynck2006thea,chinesta2010recent,ammar2010convergence,ammar2012proper} and stochastic \cite{nouy2008generalized,nouy2009recent,nouy2010proper} approaches where rather than PGD, the denomination \emph{Generalized Spectral Decomposition} (GSD) is used.
In the multiparametric nonlinear context, the PGD as solver within the LATIN framework was successfully used for dealing with a wide range of problems including, heat transfer \cite{heyberger2012multiparametric,heyberger2013rational}, elasto-visco-plasticity \cite{relun2013model,neron2015time}, and more recently reinforced concrete \cite{vitse2019dealing}.
The efficiency of the PGD made it possible to deal with problems involving upto a dozen of parameters but passed that number, the PGD seem to loose its empirically observed convergence properties. Recent developments \cite{ladeveze2018extended,paillet2018door}, based on the Saint-Venant principle, propose to overcome this limitation and extend the scope of the method to the case of high dimension.

The ambition of this work is to propose an extension of the LATIN-PGD framework to the problems raised in nonlinear dynamics by earthquake engineering applications. The applicability of the PGD for solving transient dynamic problems is debatable; one can indeed suspect that a low-rank approximation with separated variables modes will not be the most suitable representation for approaching a propagative type solution. Some conclusions in this sense have already been drawn by other authors \cite{boucinha2013space,boucinha2014ideal} when using a PGD-based solver for wave-propagation problems in an elastic media.
However, the frequency domain concerned by earthquake engineering applications is much more reduced and the solutions to be approximated by the PGD in this work must be seen as those of a forced vibration problem. In other words, there is no reason for the PGD to be ineffective where classical modal truncation gives good results. While mentioning forced vibration problems, let us also quote reference \cite{chevreuil2012model} where the PGD is employed in the frequency domain for computing parameterized FRFs. Another novelty introduced in the present paper is the use of the Time Discontinuous Galerkin Method (TDGM) to obtain the time functions of the PGD decomposition incrementally over the whole time interval. This enables an efficient computation of the time functions since the discretized operators to be inverted in each discontinuous time interval are of reduced and constant size. This property is of particular interest in differents fields, for instance seismic engineering applications, where the excitation could be of relatively long duration, or the treatment of fatigue problems. The idea of using TDGM in the LATIN-PGD solver is not new, in fact it has been extensively used in previous works \cite{nouy2003strategie,neron2004strategie,gupta2005mesodynamique,passieux2008approximation,nachar2019optimisation}, however, in those references, the TDGM was not applied to incrementally solve the temporal PGD functions while minimizing the complicated functionals that arise in the LATIN-PGD formulation due to numerical difficulties and stability issues \cite{passieux2008approximation}.

The paper is organized as follows. Section \ref{sec:ref_prob} introduce the governing equations and gives details on the concrete's constitutive relations used in this paper. Section \ref{sec:LATIN_sol} gives an overall presentation of the LATIN-PGD methodology. Section \ref{sec:Implementation} in turn, gives implementation details on the local and global stages whose alternation is central to the approach. In addition, the incremental resolution strategy in time using Time-Discontinuous Galerkin method is presented. Section \ref{sec:Num_Results} proposes an illustration of the results that can be obtained using a simple concrete beam submitted to imposed displacements at support. Finally, Section \ref{sec:concl_persp} provides some conclusions.

\newpage

\section{Equations of the reference problem}\label{sec:ref_prob}

\begin{figure}[!ht]
\centering
\vspace{0pt}
\begin{tikzpicture}[line cap=round,line join=round,>=stealth,scale=1.0]
\node[anchor=south west, inner sep=0pt, outer sep=0pt] at (0,0) {\includegraphics[width=2.5cm]{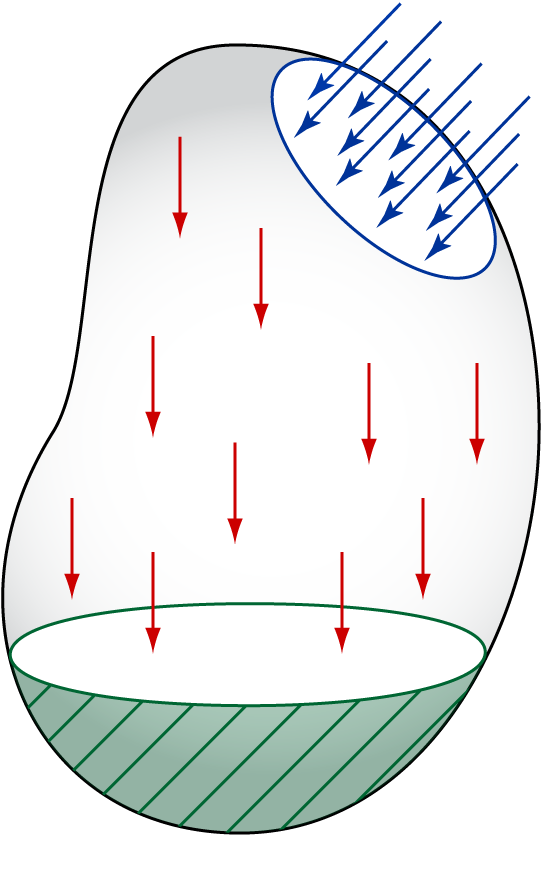}} ;
\draw (0.28,2.3) node [left] {\color{black}$\DS$} ;
\draw (2.5,4.2) node [] {\color{qqttzz}$\vect{f}^{N}$} ;
\draw (2.3,0.5) node [] {\color{qqwwtt}$\vect{u}^{D}$} ;
\draw (1.25,2.2) node [] {$\rho \vect{f}$} ;
\draw [->][black][line width=1pt][cap=round] (0,0) -- (3.0,0);
\draw (0,0) node[] {\color{black}$\scriptscriptstyle{|}$} ;
\draw (0,0) node[below] {\color{black}$0$} ;
\draw (2.6,0) node[] {\color{black}$\scriptscriptstyle{|}$} ;
\draw (2.6,0) node[below] {\color{black}$T$} ;
\draw (1.4,0) node[below] {\color{black}$\DT$} ;
\end{tikzpicture}
\caption{The mechanical domain under study.}
\label{fig:ref_prob}
\end{figure}
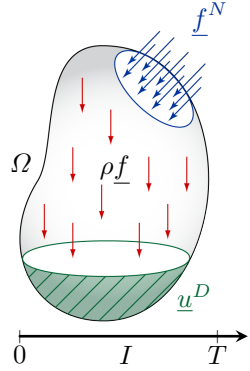

Let us consider the medium of Figure \ref{fig:ref_prob} occupying the domain $\DS \subset \R^d$ with $d \in \{1,2,3\}$, on a time domain $\DT=[0,T]$ and with constant boundary $\partial \DS = \partial_{N}\DS \oplus \partial_{D}\DS $ over time, where $\partial_{N}\DS$ and $\partial_{D}\DS$ are the boundaries related to the imposed Neumann and Dirichlet conditions respectively.
This structure is submitted to surface forces $\vect{f}^{N}$ on $ \partial_{N}\DS \times \DT$ (Neumann boundary conditions), to imposed displacements $\vect{u}^{D}$ on $\partial_{D}\DS \times \DT$ (Dirichlet boundary condition) and to volumetric forces $\rho \vect{f}$ on $\DS \times \DT$.
%
%

\subsection{The reference problem}\label{SEC:2:REF:Admissibility}

\begin{probleme}
\label{PROB:SF}
The reference problem consists in finding a displacement field $\vect{u}(\vx,t) \in \Su$ and a stress field $\ctens{\sigma}(\vx,t) \in \Esig$ verifying:
\begin{subequations}
\begin{enumerate}[label=$\bullet$,leftmargin=10pt]
  \item initial conditions:
\end{enumerate}
  \begin{equation}
    \label{EQ:REF:CLim}
     \tableq{\mbox{on} \ \DS,}
     {\begin{array}{c} \vect{u} |_{\substack{t=0^{+}}}= \vect{u}_{\init} \\
      \dot{\vect{u}} |_{\substack{t=0^{+}}}= \dot{\vect{u}}_{\init}
     \end{array}}
  \end{equation}
\begin{enumerate}[label=$\bullet$,leftmargin=10pt]
  \item dynamic equilibrium equation:
\end{enumerate}
  \begin{equation}
    \label{EQ:REF:Eq}
    \tableq{\mbox{on} \ \DS\times\DT,}
    {\rho \ddot{\vect{u}} = \mbox{div}(\ctens{\sigma}) + \rho \vect{f}}
  \end{equation}
\begin{enumerate}[label=$\bullet$,leftmargin=10pt]
  \item Neumann and Dirichlet boundary conditions:
\end{enumerate}
  \begin{align}
    \label{EQ:REF:Neumann}
    \tableq{\mbox{on} \ \partial_{N}\DS \times \DT,}
    {\ctens{\sigma} \dotp \vect{n} = \vect{f}^{N}} \\
    \label{EQ:REF:Dirichlet}
     \tableq{\mbox{on} \ \partial_{D}\DS \times \DT,}
     {\vect{u}=\vect{u}^{D}}
  \end{align}
\begin{enumerate}[label=$\bullet$,leftmargin=10pt]

  \item Constitutive equations relating the stress tensor $\ctens{\sigma}$ to the strain tensor $\ctens{\varepsilon}(\vect{u}) = \frac{1}{2} \left( \nabla \vect{u} + \nabla \vect{u}^{T} \right)$, which can be read in a general manner under the form:
\end{enumerate}
  \begin{equation}
    \label{EQ:REF:RdC}
    \tableq{\mbox{on} \ \DS\times\DT,}
    {\Prth{\ctens{\sigma},\IV} \eq \RdC \Prth{ \ctens{\dot{\varepsilon}}(\vect{u})\big|_{\tau} , \dot{\IV}\big|_{\tau} , \tau \leq t}}
  \end{equation}
\begin{enumerate}[label={},leftmargin=10pt]
  \item where $\IV$ contains the different primal and dual internal variables needed to describe the behavior of the domain. The details related to this behavior are presented in the following Section \ref{sec:iso_damage_cr}.
\end{enumerate}
\end{subequations}
\end{probleme}

\

\subsection{Isotropic damage model for concrete materials}\label{sec:iso_damage_cr}

The following sub-sections present the constitutive relations considered to model damage evolution on a concrete medium. In the earthquake engineering context that is considered in this work, alternate cyclic loading is expected on the solid domain. Thus, two important
characteristics of concrete must be reproduced by the model namely: $(i)$ the asymmetric
traction-compression behavior referred to as \emph{unilateral effect} \cite{mazars1990unilateral} and $(ii)$ the crack-reclosure
phenomenon. The simplified concrete modeling, developed in \cite{vitse2019dealing} from previous work by \cite{richard2013continuum}
and \cite{vassaux2015regularised}, is used here. Among other simplifications, no damage in compression is considered.


The quasi-brittle behavior of the concrete medium in tension is modeled in the classical
continuum damage mechanics framework \cite{lemaitre2005engineering}  using a scalar damage variable denoted $d$. In
tension, the effective stress tensor in the concrete medium classically writes (see Figure \ref{FIG:MAT:Concrete:a}):
\begin{equation}
\tilde{\ctens{\sigma}}_{\med} = \frac{\ctens{\sigma}_{\med}}{1-d}
\end{equation}

In compression however (see Figure \ref{FIG:MAT:Concrete:b}), the main idea proposed in \cite{vassaux2015regularised} for handling unilateral effect and progressive micro-cracks re-closure phenomena consists in the use of an additional stress tensor $\ctens{\sigma}_{\cra}$. The stress tensor in the concrete Representative Volume Element (RVE) then writes using a Kelvin-Voigt description (see Figure \ref{fig:concrete_stress_sep}):
\begin{equation}
\ctens{\sigma} = \ctens{\sigma}_{\med} + \ctens{\sigma}_{\cra}
\end{equation}
introducing $\ctens{\sigma}_{\med}$ the stress in the concrete medium and $\ctens{\sigma}_{\cra}$ accounting for the proportion of closed micro-cracks.

An effective stress tensor associated to gradually closed micro-cracks can then be defined as (see Figure \ref{FIG:MAT:Concrete:b} and \ref{FIG:MAT:Concrete:c}.):
\begin{equation}
\tilde{\ctens{\sigma}}_{\cra} = \frac{\ctens{\sigma}_{\cra}}{d}
\end{equation}
Still following the Kelvin-Voigt description and using coherent indexing, the strain tensor
writes:
\begin{equation}
\ctens{\varepsilon} = \ctens{\varepsilon}_{\med} = \ctens{\varepsilon}_{\cra}
\end{equation}


\def\asc{0.8} 	 
\def\hVs{0.5cm} 
\def\hMil{0.53} 
\def\hVse{0.8cm} 
\def\tW{\textwidth}

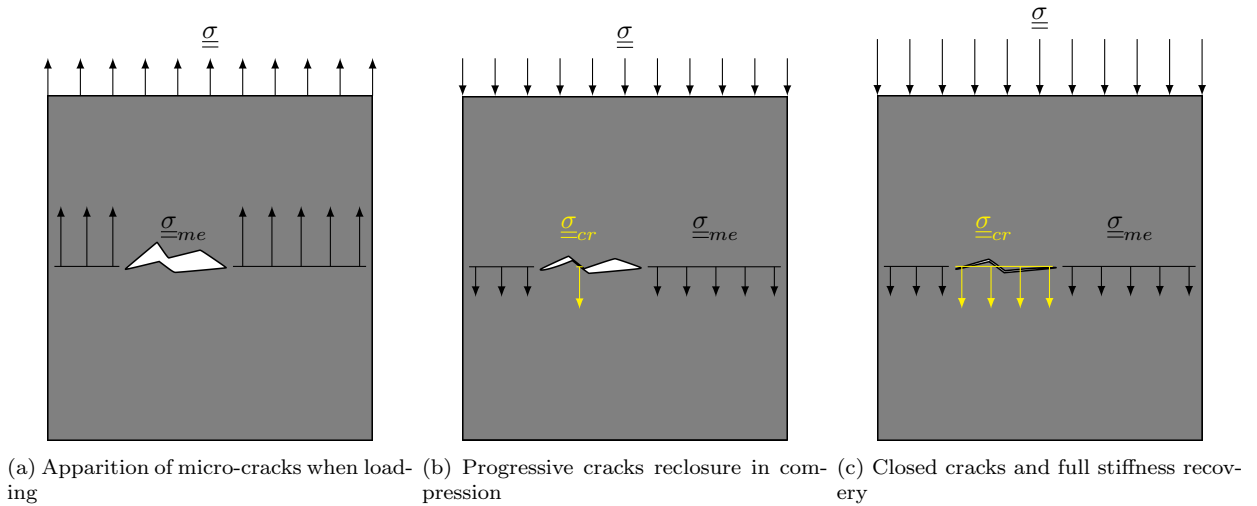
\begin{figure}[H]
\centering
\begin{subfigure}[t]{0.3\textwidth}
\centering
\begin{tikzpicture}[scale=1]
\node[anchor=north west,inner sep=0pt,outer sep=0pt] at (0,0) {\includegraphics[width=\asc\textwidth]{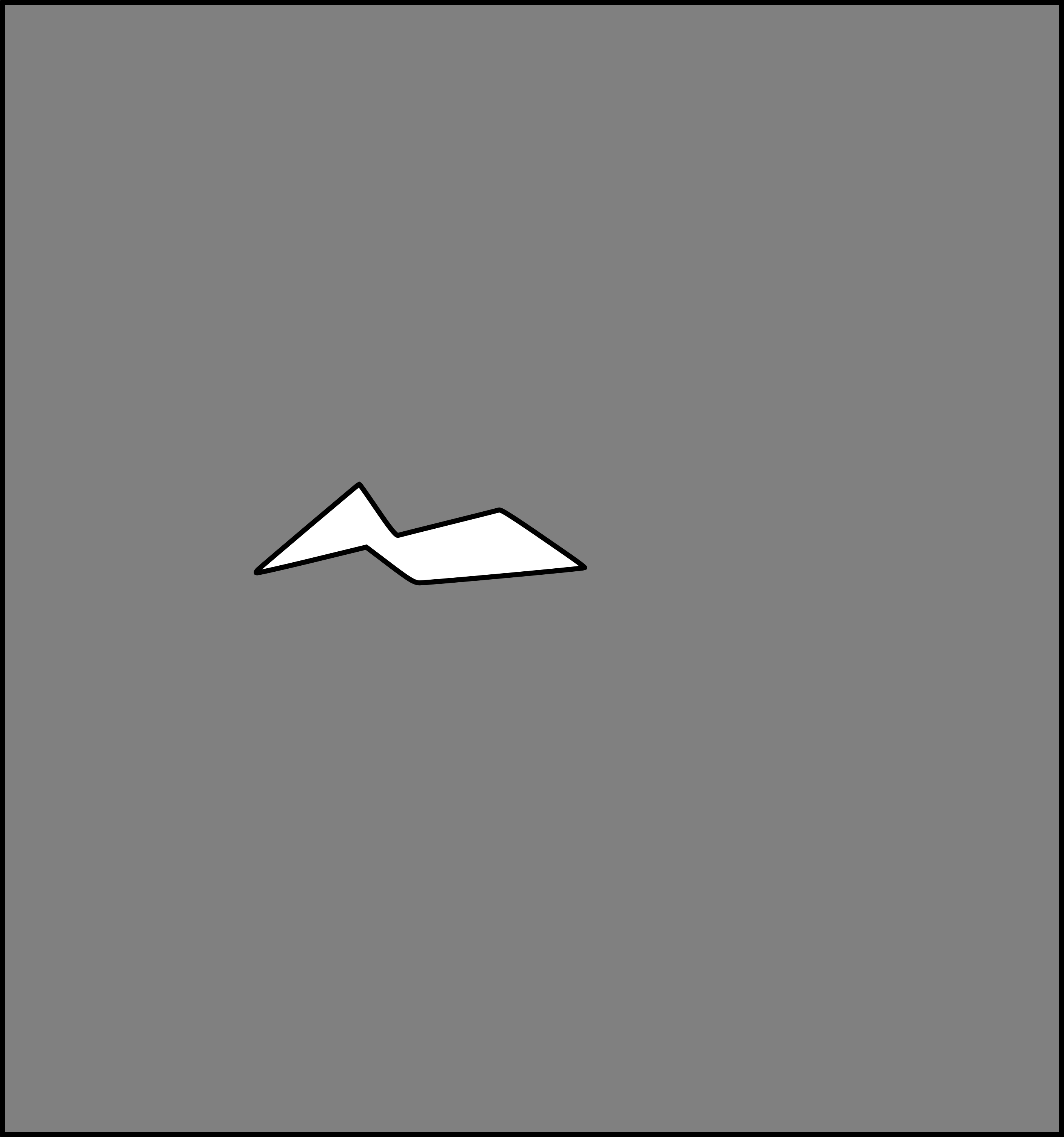}};

\foreach \sA in {0.0,0.1,0.2,0.3,0.4,0.5,0.6,0.7,0.8,0.9,1.0}{
\draw[->,>=latex] (\sA*\asc*\textwidth,0) to (\sA*\asc*\textwidth,\hVs) ;
}
\draw (0.5*\asc*\textwidth,\hVs) node[above] {\color{black}$\ctens{\sigma}$};

\draw[-,color=black] (0.02*\asc*\textwidth,-\hMil*\asc*\textwidth) to (0.22*\asc*\textwidth,-\hMil*\asc*\textwidth) ;
\draw[-,color=black] (0.57*\asc*\textwidth,-\hMil*\asc*\textwidth) to (0.98*\asc*\textwidth,-\hMil*\asc*\textwidth) ;

\draw (0.42*\asc*\textwidth,-0.9*\hMil*\asc*\textwidth) node[above] {\color{black}$\ctens{\sigma}_{\med}$};

\foreach \sA in {0.04,0.12,0.20}{
\draw[->,>=latex,color=black] (\sA*\asc*\textwidth,-\hMil*\asc*\tW) to (\sA*\asc*\textwidth,-\hMil*\asc*\textwidth+\hVse) ;
}

\foreach \sA in {0.6,0.69,0.78,0.87,0.96}{
\draw[->,>=latex,color=black] (\sA*\asc*\textwidth,-\hMil*\asc*\textwidth) to (\sA*\asc*\textwidth,-\hMil*\asc*\textwidth+\hVse) ;
}
\end{tikzpicture}
\caption{Apparition of micro-cracks when loading}
\label{FIG:MAT:Concrete:a}
\end{subfigure}
\begin{subfigure}[t]{0.3\textwidth}
\centering
\begin{tikzpicture}[scale=1]
\node[anchor=north west,inner sep=0pt,outer sep=0pt] at (0,0) {\includegraphics[width=\asc\textwidth]{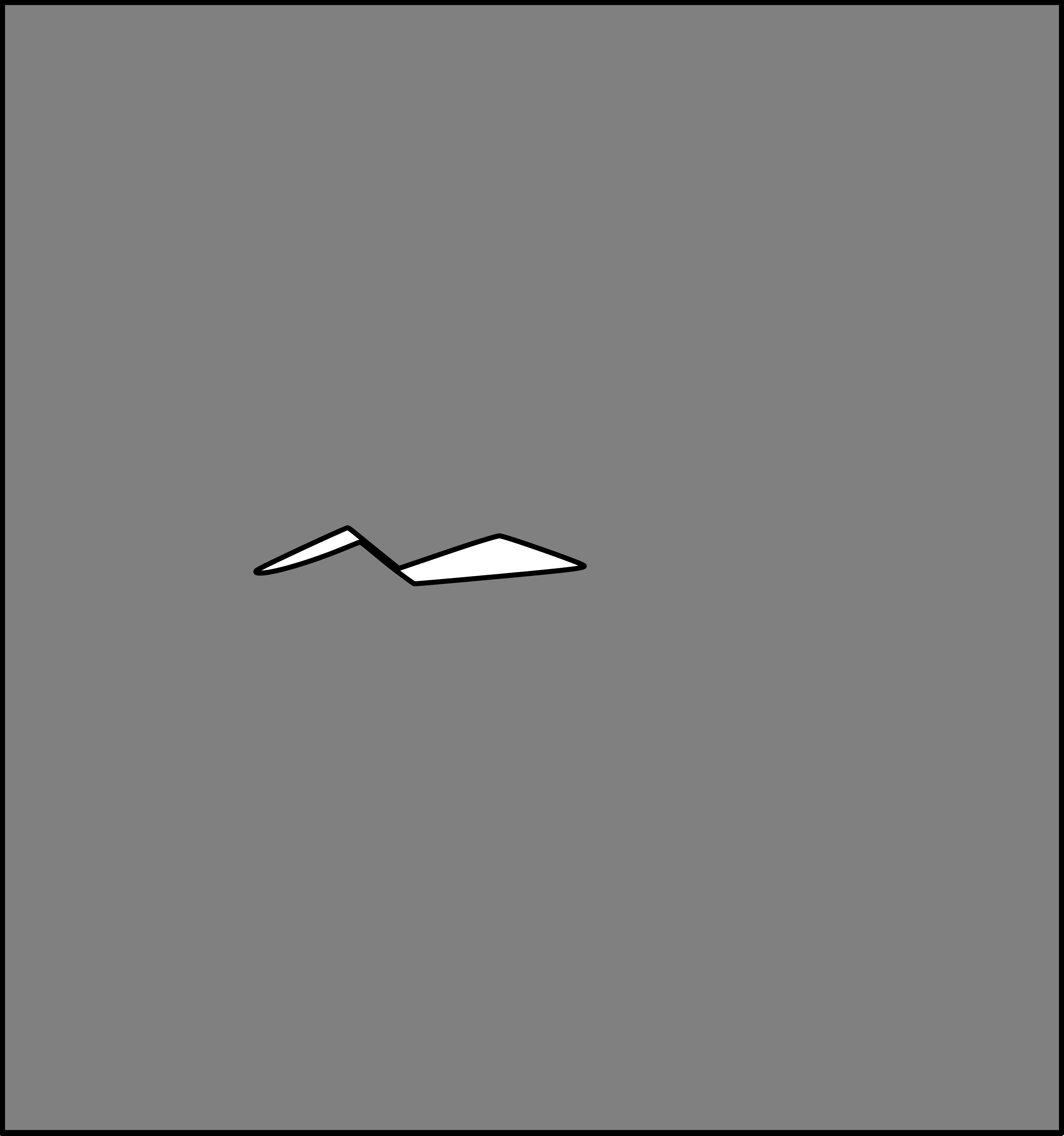}} ;

\foreach \sA in {0.0,0.1,0.2,0.3,0.4,0.5,0.6,0.7,0.8,0.9,1.0}{
\draw[->,>=latex] (\sA*\asc*\textwidth,\hVs) to (\sA*\asc*\textwidth,0) ;
}
\draw (0.5*\asc*\textwidth,\hVs) node[above] {\color{black}$\ctens{\sigma}$};

\draw[-,color=black] (0.02*\asc*\textwidth,-\hMil*\asc*\textwidth) to (0.22*\asc*\textwidth,-\hMil*\asc*\textwidth) ;
\draw[-,color=black] (0.57*\asc*\textwidth,-\hMil*\asc*\textwidth) to (0.98*\asc*\textwidth,-\hMil*\asc*\textwidth) ;

\draw (0.78*\asc*\textwidth,-0.9*\hMil*\asc*\textwidth) node[above] {\color{black}$\ctens{\sigma}_{\med}$};

\foreach \sA in {0.04,0.12,0.20}{
\draw[->,>=latex,color=black] (\sA*\asc*\textwidth,-\hMil*\asc*\textwidth) to (\sA*\asc*\textwidth,-\hMil*\asc*\textwidth-0.5*\hVse) ;
}

\foreach \sA in {0.6,0.69,0.78,0.87,0.96}{
\draw[->,>=latex,color=black] (\sA*\asc*\textwidth,-\hMil*\asc*\textwidth) to (\sA*\asc*\textwidth,-\hMil*\asc*\textwidth-0.5*\hVse) ;
}

\draw[-,color=yellow] (0.35*\asc*\textwidth,-\hMil*\asc*\textwidth) to (0.37*\asc*\textwidth,-\hMil*\asc*\textwidth) ;

\foreach \sA in {0.36}{
\draw[->,>=latex,color=yellow] (\sA*\asc*\textwidth,-\hMil*\asc*\textwidth) to (\sA*\asc*\textwidth,-\hMil*\asc*\textwidth-0.7*\hVse) ;
}
\draw (0.36*\asc*\textwidth,-0.9*\hMil*\asc*\textwidth) node[above] {\color{yellow}$\ctens{\sigma}_{\cra}$};

\end{tikzpicture}
\caption{Progressive cracks reclosure in compression}
\label{FIG:MAT:Concrete:b}
\end{subfigure}
\begin{subfigure}[t]{0.3\textwidth}
\centering
\begin{tikzpicture}[scale=1]
\node[anchor=north west,inner sep=0pt,outer sep=0pt] at (0,0) {\includegraphics[width=\asc\textwidth]{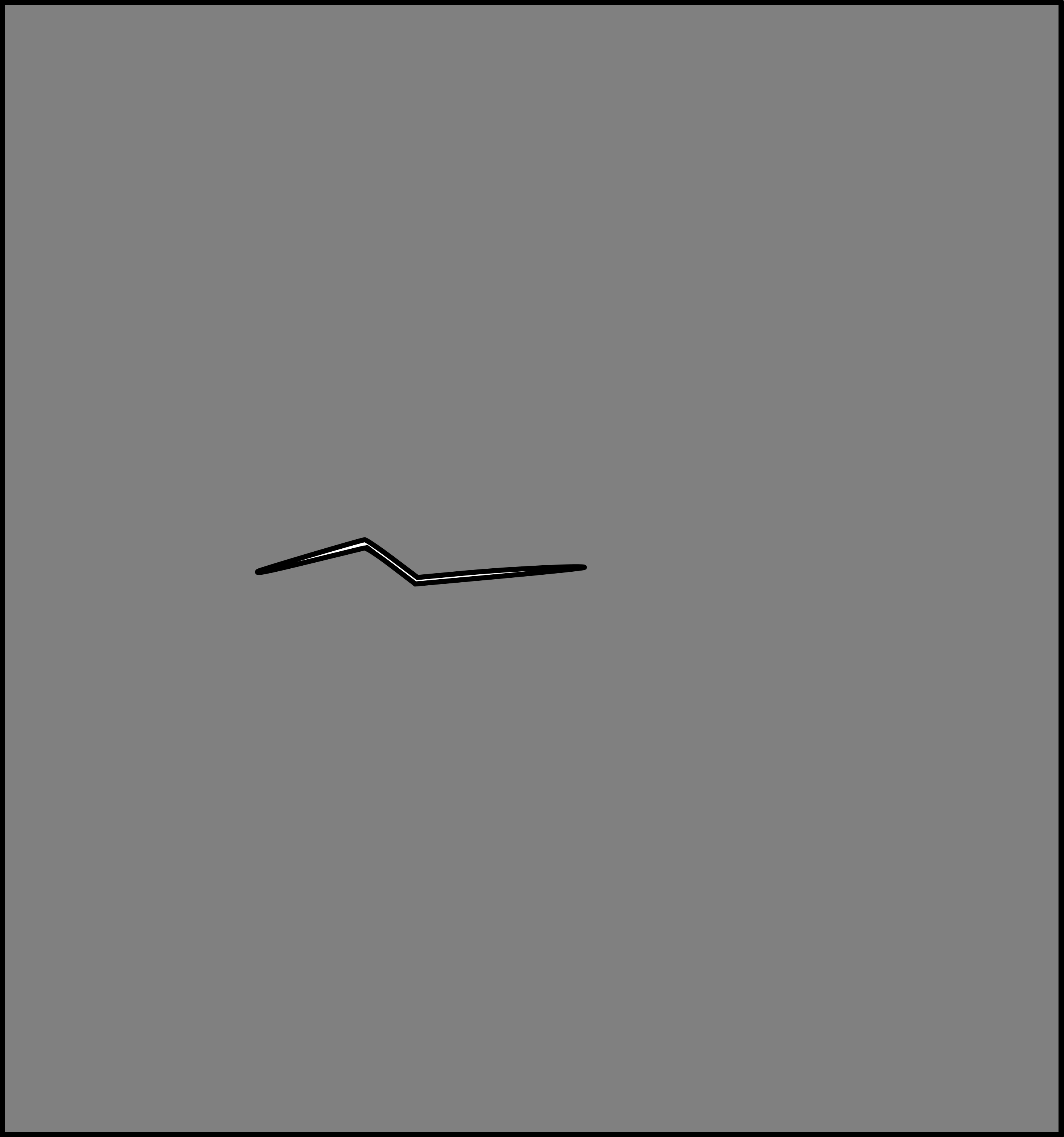}} ;

\foreach \sA in {0.0,0.1,0.2,0.3,0.4,0.5,0.6,0.7,0.8,0.9,1.0}{
\draw[->,>=latex] (\sA*\asc*\textwidth,1.5*\hVs) to (\sA*\asc*\textwidth,0) ;
}
\draw (0.5*\asc*\textwidth,1.5*\hVs) node[above] {\color{black}$\ctens{\sigma}$} ;

\draw[-,color=black] (0.02*\asc*\textwidth,-\hMil*\asc*\textwidth) to (0.22*\asc*\textwidth,-\hMil*\asc*\textwidth) ;
\draw[-,color=black] (0.57*\asc*\textwidth,-\hMil*\asc*\textwidth) to (0.98*\asc*\textwidth,-\hMil*\asc*\textwidth) ;

\draw (0.78*\asc*\textwidth,-0.9*\hMil*\asc*\textwidth) node[above] {\color{black}$\ctens{\sigma}_{\med}$};

\foreach \sA in {0.04,0.12,0.20}{
\draw[->,>=latex,color=black] (\sA*\asc*\textwidth,-\hMil*\asc*\textwidth) to (\sA*\asc*\textwidth,-\hMil*\asc*\textwidth-0.5*\hVse) ;
}

\foreach \sA in {0.6,0.69,0.78,0.87,0.96}{
\draw[->,>=latex,color=black] (\sA*\asc*\textwidth,-\hMil*\asc*\textwidth) to (\sA*\asc*\textwidth,-\hMil*\asc*\textwidth-0.5*\hVse) ;
}

\draw[-,color=yellow] (0.24*\asc*\textwidth,-\hMil*\asc*\textwidth) to (0.54*\asc*\textwidth,-\hMil*\asc*\textwidth) ;

\foreach \sA in {0.26,0.35,0.44,0.53}{
\draw[->,>=latex,color=yellow] (\sA*\asc*\textwidth,-\hMil*\asc*\textwidth) to (\sA*\asc*\textwidth,-\hMil*\asc*\textwidth-0.7*\hVse) ;
}
\draw (0.36*\asc*\textwidth,-0.9*\hMil*\asc*\textwidth) node[above] {\color{yellow}$\ctens{\sigma}_{\cra}$};

\end{tikzpicture}
\caption{Closed cracks and full stiffness recovery}
\label{FIG:MAT:Concrete:c}
\end{subfigure}
\caption{Handling unilateral effect in concrete RVE -- uniaxial illustration with only one micro-crack.}
\label{fig:concrete_stress_sep}
\end{figure}
Section \ref{sec:QS_behavior} describes how the quasi-brittle behavior in tension is modeled, assuming that $\ctens{\sigma}_{\cra} = 0$. In turn Section \ref{sec:re_closure_crack} describes how tensor $\ctens{\sigma}_{\cra}$ is added to $\ctens{\sigma}_{\med}$  in a progressive manner for compression load phases. Again, no damage in compression will be taken into account in the modeling. The scalar variable $d$ only models the micro-cracks
occurring during tension loading phases.

More sophisticated modeling strategies can be taken into account to model the unilateral effect and can be used in conjunction with the LATIN-PGD approach without difficulties. For instance, the anisotropic damage (tensorial damage variable, handling
damage in tension and compression) presented in \cite{souid2009pseudodynamic} and suitable for alternative low-dynamic loading could have also been considered.

\subsubsection{Quasi-brittle behavior in tension}\label{sec:QS_behavior}

According to the continuum damage mechanics framework \cite{lemaitre1994mechanics}, the Helmoltz free enthalpy, taken as the state potential of the cracked material, is a function of all the primal
state variables and writes:
\begin{equation}
\mathit{\Psi}^{\med}(d,\ctens{\varepsilon},z) = \frac{1}{2}(1-d)\ \ctens{\varepsilon} : \HookeC : \ctens{\varepsilon} \ + \ H(z)
\end{equation}
introducing the classical fourth order Hooke's tensor $\HookeC$ (undamaged configuration) and the potential $H(z)$ and variable $z$ associated to isotropic softening related to damage. The choice of the potential function
$H(z)$ governs the post-peak behavior of the concrete medium in tension. In \cite{richard2013continuum}, the following
choice is made:
\begin{equation}
H(z) = \frac{1}{A_{d}} \left( -z +\text{log}(1+z) \right)
\end{equation}
The following variables are then introduced by duality writing:
\begin{equation}
Z = \frac{\partial \mathit{\Psi}^{\med}}{\partial z}  \quad \text{and} \quad  Y = -\frac{\partial \mathit{\Psi}^{\med}}{\partial d}
\end{equation}
$Y$ is an energy release rate; the dissipated energy released in the damage process writes $Y \dot{d}$.
Since damage mainly occurs in tension, this energy rate is defined as $Y = \frac{1}{2} \ \langle \ctens{\varepsilon} \rangle_{+} : \HookeC : \langle \ctens{\varepsilon} \rangle_{+}$, following the classical Mazars description \cite{mazars1984application},
where the Macaulay brackets $\langle \square \rangle_{+}$  denote the positive part of the tensor $\square$.
$Z$ in turn, also homogeneous to an energy rate, is the thermodynamical force involved in the definition of the damage threshold and defines the actual size of the elastic domain. Thus, the surface threshold defined in terms of energy released through damage reads:
\begin{equation}
f(Y,Z;Y_{0}) = Y - (Y_{0} + Z) \qquad  \begin{cases}
f<0 \qquad \text{elasticity} \\
f \geqslant 0 \qquad \text{damage activation}
\end{cases}
\end{equation}
An analytical expression of damage can be obtained \cite{richard2010isotropic}, after assuming associative flow rule, writing normality rules and consistency conditions, so that the evolution laws become:
\begin{equation}\label{eq:damage_model}
d = 1 - \frac{1}{1 + A_{d}(Y - Y_{0})} \qquad \text{and} \qquad z = -d
\end{equation}
Classical continuum damage theory based models may have difficulties to describe fracture properly; indeed, the loss of ellipticity is a recurrent issue when dealing with materials with softening behavior, leading to a localization of the deformation and mesh dependency
issues \cite{bazant1984continuum}. One way of overcoming such numerical difficulties is to use localization limiters \cite{lasry1988localization}. In this work the damage is regularized using a viscosity law. 
A strategy similar to what is described in \cite{allix1997delayed,allix2003delay,allix2013bounded} is implemented. 
The introduction of a delay effect leads to a new damage evolution law that can be written as:
\begin{equation}\label{eq:Damage_delay}
\dot{d} = \frac{1}{\tau_{c}} \CRO{ 1 - \text{exp}\left( -a \left[ \langle \bar{d}(Y) - d \rangle_{+} \right]   \right) }
\end{equation}
where $\bar{d}(Y)$ is the damage value computed using equation \eqref{eq:damage_model} and the scalars $\left( \tau_{c}, a \right)$ are two new parameters associated to damage regularization. 
The damage variable $d$ now verifies a nonlinear first order differential equation. 
The maximum damage rate is given by $\dot{d}_{max} = \frac{1}{\tau_{c}}$ and the more or less brittle character of the damage evolution law is governed by the choice of the parameter $a$.

\subsubsection{Progressive micro-crack re-closure}\label{sec:re_closure_crack}

The unilateral effect is modelled using an auxiliary stress tensor $\ctens{\sigma}_{c}$ accounting for the progressive micro-cracks re-closure in compression and enabling full stiffness recovery. In this manner, even though permanently damaged in tension, the concrete medium will behave in compression almost independently from its history in tension. The progressive stiffness regain can be handled by the means of a diffeomorphism $\mathcal{F}:\mathbb{R}^{3} \otimes \mathbb{R}^{3} \longrightarrow \mathbb{R}^{3} \otimes \mathbb{R}^{3}$ such that:
\begin{equation}
\tilde{\ctens{\sigma}}_{\cra} = \HookeC : \mathcal{F}(\ctens{\varepsilon})
\end{equation}
and with $\mathcal{F}(\ctens{\varepsilon})$ verifying:
\begin{equation}\label{eq:cond_eps_concrete}
\begin{cases}
\mathcal{F}(\ctens{\varepsilon}) \sim \ctens{\varepsilon}  \qquad \text{in compression} \\
\mathcal{F}(\ctens{\varepsilon}) \sim 0 \qquad \text{in tension}
\end{cases}
\end{equation}
The tangent behavior is then defined writing:
\begin{equation}
d  \tilde{\ctens{\sigma}}_{\cra} = \HookeC : \left[ \frac{\partial \mathcal{F}}{\partial \ctens{\varepsilon}}  \right] : d  \ctens{\varepsilon}
\end{equation}
where $\left[ \frac{\partial \mathcal{F}}{\partial \ctens{\varepsilon}}  \right]$ is a fourth order tensor.
In this work, following the lines of \cite{vassaux2015regularised}, the following function is used:
\begin{equation}\label{eq:closure_relation_eps}
\tilde{\ctens{\sigma}}_{\cra} = \HookeC : \left[  
\ctens{\varepsilon} -   \left( \frac{\ctens{\varepsilon}_{\text{max}}}{a_{c}}   \right) \text{log} \left( 1 + \text{exp}\left(  a_{c} \frac{\text{tr}(\ctens{\varepsilon})}{\text{tr}(\ctens{\varepsilon}_{\text{max}})}  \right) \right)
\right]
\end{equation}
where $\ctens{\varepsilon}_{\text{max}} = \ctens{\varepsilon}(t_m)$ with $\displaystyle t_m = \arg \max_{0<\tau<t} \text{tr}(\ctens{\varepsilon}(\tau))$
and  where $a_{c}$ is a constant parameter that can be modified to define a more or less progressive crack-closure ($a_{c} \to \infty $ for more sudden crack-closure). 
This expression for $\tilde{\ctens{\sigma}}_{\cra}$ verifies condition \eqref{eq:cond_eps_concrete} and thus by writing $\ctens{\sigma}_{\cra} = d \ \tilde{\ctens{\sigma}}_{\cra}$, 
one can prove that the stiffness is fully recovered in compression. 
More details and uni-axial interpretations of the different parameters involved in the crack-closure relation \eqref{eq:closure_relation_eps} can be found in the initial reference \cite{vassaux2015regularised}.

		\subsubsection{Uniaxial response of the concrete medium -- Comparison with test results}
		\label{SEC:2:MAT:Concrete:Comparison}

\begin{figure}[!ht]
\centering
\begin{subfigure}[c]{0.45\textwidth}
\includegraphics[width=\textwidth]
{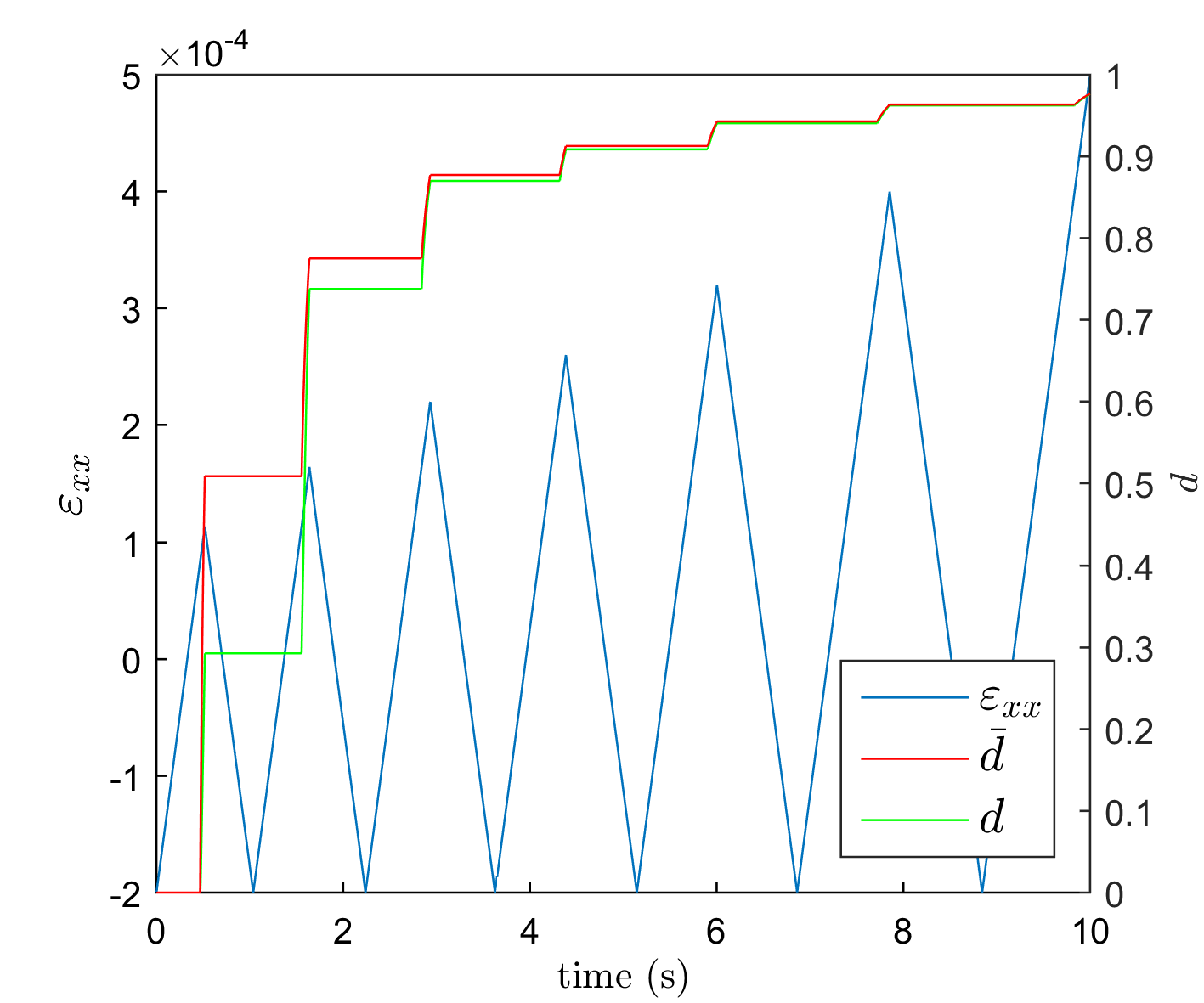}
\end{subfigure}
\begin{subfigure}[c]{0.45\textwidth}
\includegraphics[width=\textwidth]
{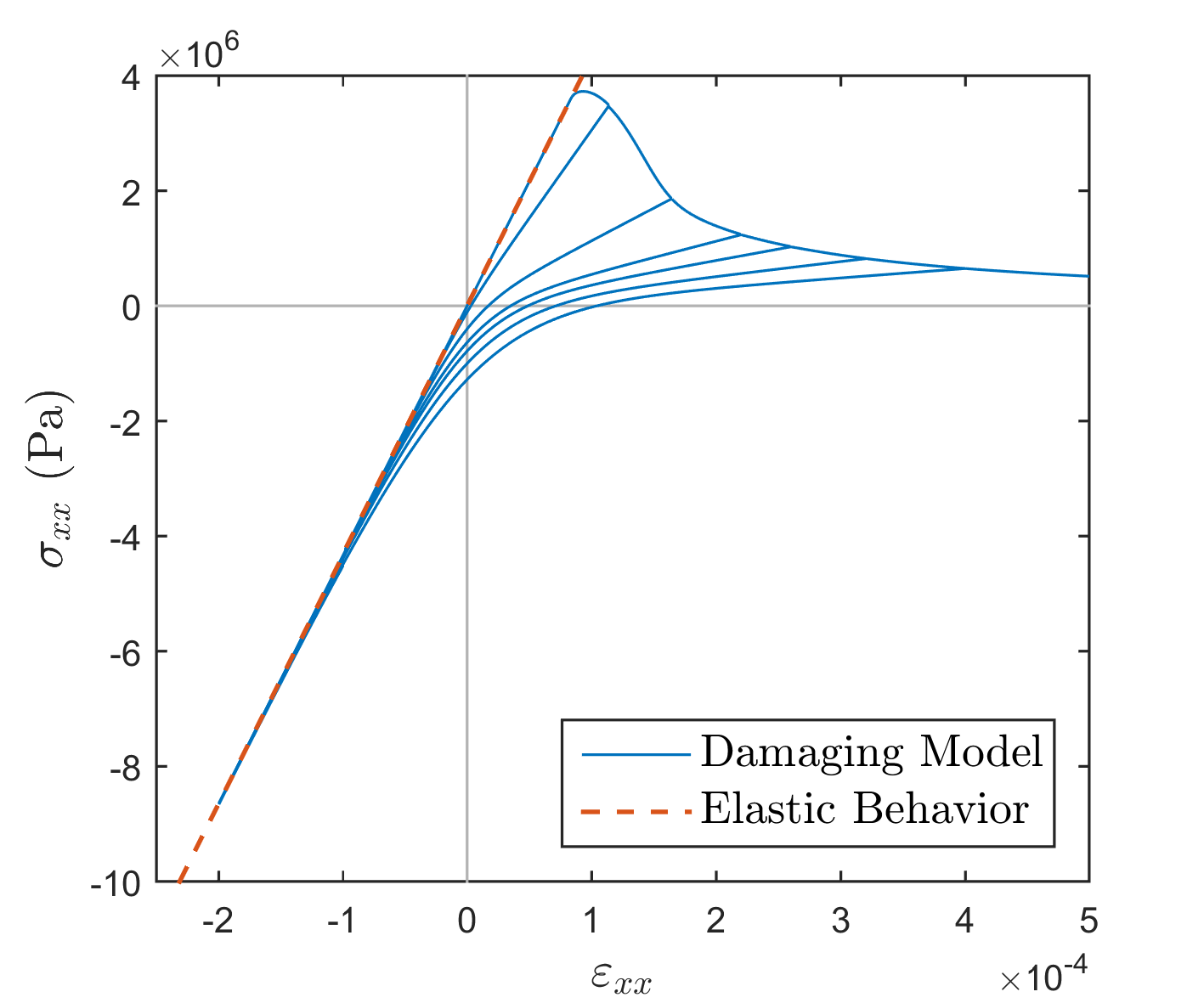}
\end{subfigure}
\caption{Mechanical response of the model at a Gauss point to a uni-axial tension-compression loading.}
\label{fig:numeric_test_beton_0D}
\end{figure}

Figure \ref{fig:numeric_test_beton_0D} shows a first illustration of the model response in a Gauss point to an uni-axial strain imposed in the $x$ direction (seven loops).
One can particularly appreciate how the softening behavior is handled by the potential $H$ and observe the evolution of damage taking into account (variable $d$) or not (variable $\bar{d}$) the damage delay eq. \eqref{eq:Damage_delay}.

\begin{figure}[!ht]
\centering
\begin{subfigure}[c]{0.45\textwidth}
\includegraphics[width=\textwidth]
{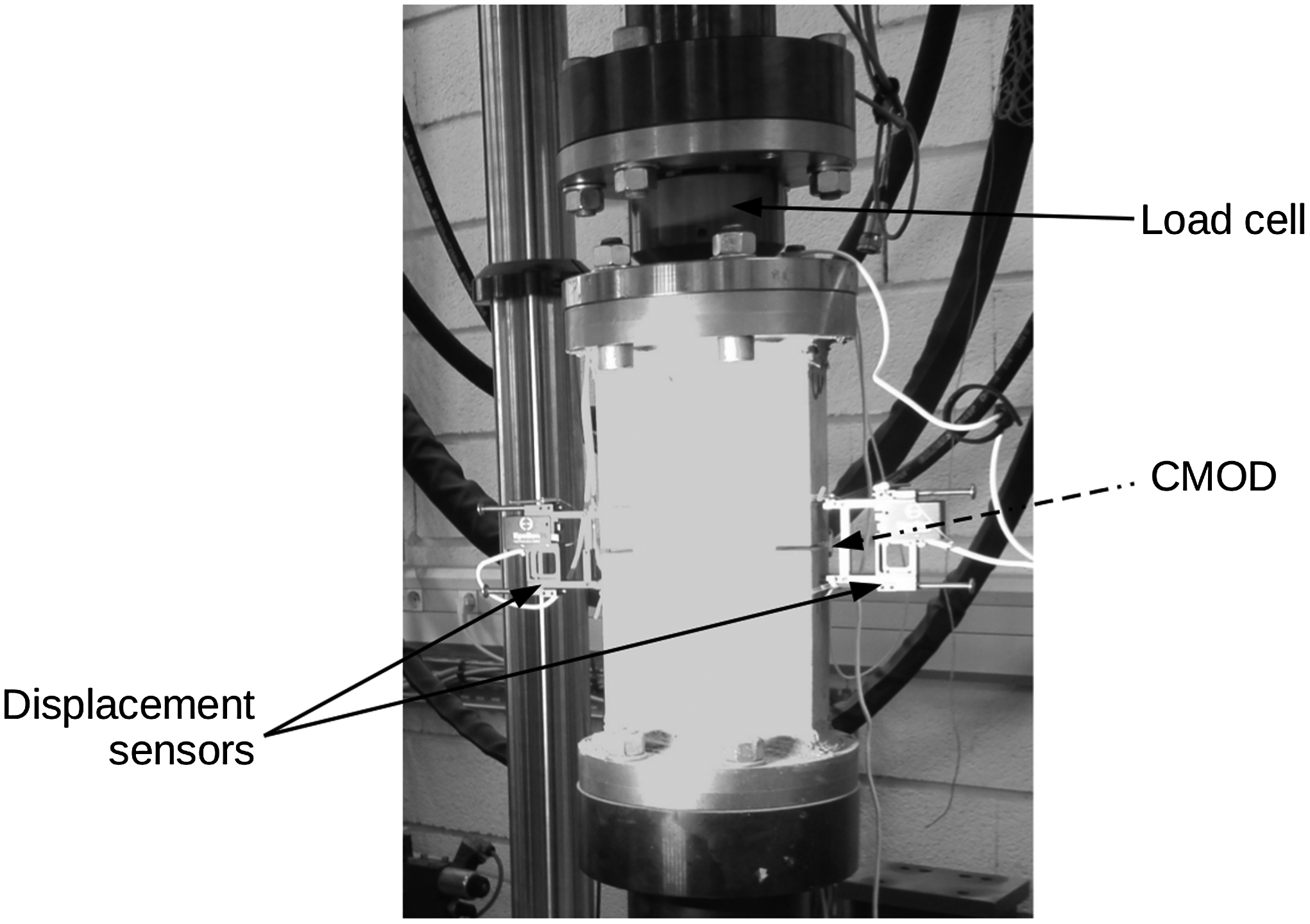}
\end{subfigure}
\begin{subfigure}[c]{0.45\textwidth}
\includegraphics[width=\textwidth]
{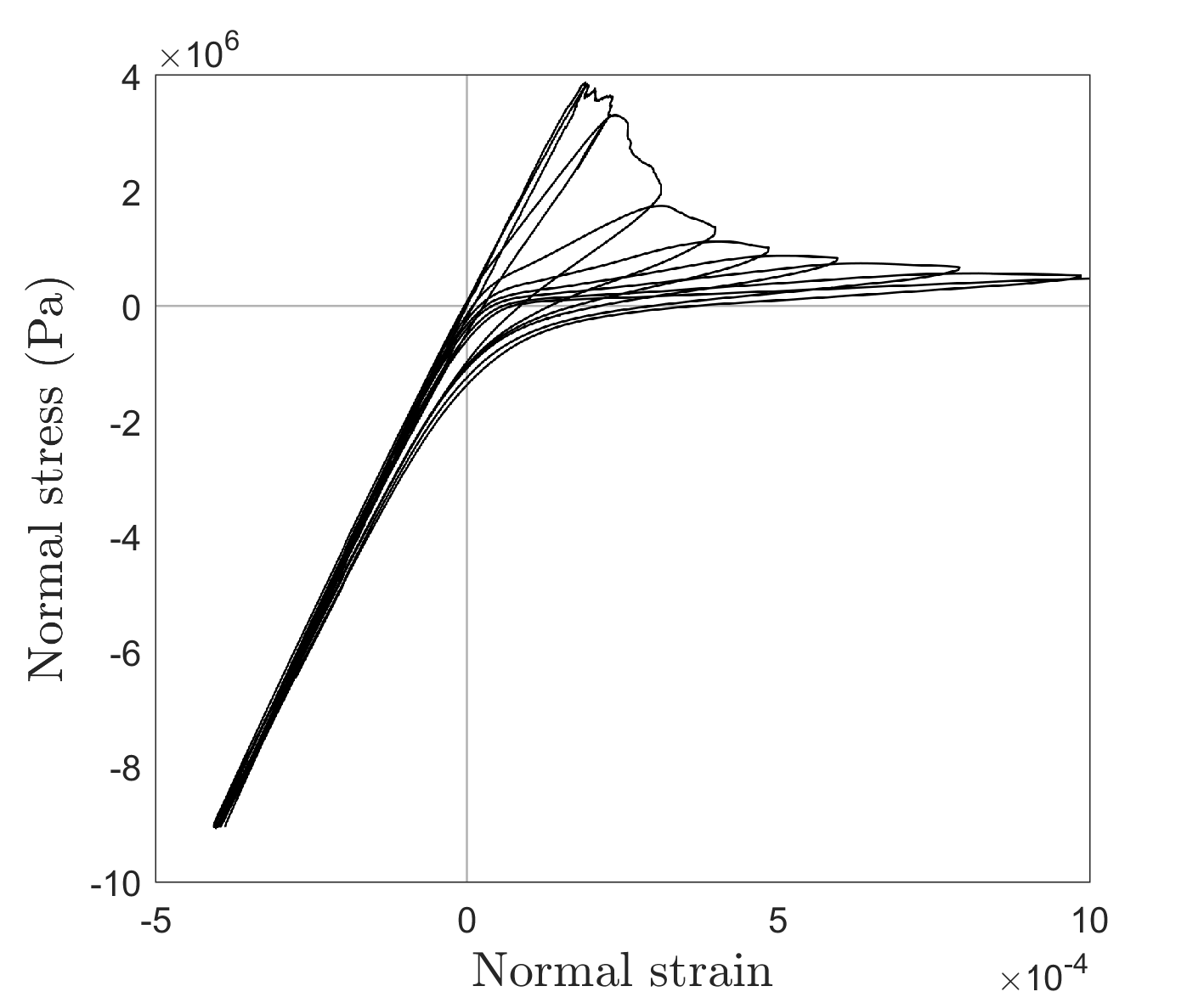}
\end{subfigure}
\caption{Uni-axial test results on double-notched concrete specimen described in \cite{nouailletas2015experimental}.
(Left) Imposed displacement in the vertical direction, two Crack Mouth Opening Displacement (CMOD)
sensors are measuring the displacement between the two faces of each notch. (Right) Stress vs. strain plot
in the vertical direction; the ``normal strain" on abscissa is an averaged value computed from the two
CMOD sensors.}
\label{fig:real_beton_test}
\end{figure}

The reference parameters used for this first 0D calibration (see Tab. \ref{tab:concrete_mat_param}) have been chosen such as reproducing the main features of the results obtained for uni-axial tests on a double-notched concrete specimen in \cite{nouailletas2015experimental}.
Figure \ref{fig:real_beton_test} reproduces for convenience the experimental set-up description and uni-axial response of the concrete specimen extracted from the above-mentioned reference.

One can observe that the simplified concrete modeling considered in this work is capable of reproducing the softening behavior in tension and handling stiffness recovery in compression, and this, while taking advantage of the regularizing action of the damage delay technique.
No dissipative mechanism is introduced in our simplified  modeling to reproduce the hysteresis  loops (unlike what can be found in \cite{richard2013continuum,vassaux2015regularised} where a visco-plastic potential is added to the constitutive relations).
However, the dissipative phenomena missing in the model description are taken into account using a $2\%$ constant modal damping of ratio typical to what have been
experimentally assessed on concrete and reinforced-concrete specimens under seismic excitation (see e.g. \cite{charbonnel2021modal}).
As previously mentioned, no damage in compression is taken into account.

\begin{table}[H]  
\centering
\begin{tabular}{p{1cm}p{2.7cm}p{3.2cm}ll}
\hline
Mat. & \multicolumn{1}{l}{Parameters} & Ref. value & Description \\
\hline
\hline
\multirow{8}{*}{\rotatebox{90}{Concrete}}
& $\rhoc$ & $\SI{2550}{kg.m^{-3}}$ & Material density \\
 & $\Ymc$ & $\SI{37.9}{GPa}$ & Young's modulus \\
 & $\Pcc$ & $\SI{0.2}{}$ & Poisson's ratio \\
 & $\Y_{0}$ & $\SI{150}{J.m^3.kg^{-1}}$ & Damage activation energy threshold \\
 & $\Ad$ & $\SI{8.10^{-3}}{J^{-1}.m^{-3}.kg}$ & Brittleness coefficient \\
 & $\dd$ & $\SI{0.05}{s}$ & Damage delay time constant \\
 & $\ddexp$ & $\SI{15}{}$ & Damage delay exp. constant \\
  & $\cc$ & $\SI{9}{}$ & Crack-closure dimensionless constant \\
& $\xi$ & $\SI{0.02}{}$ & Constant modal damping ratio \\
\hline
\end{tabular}
\caption{Reference material parameters.}
\label{tab:concrete_mat_param}
\end{table}

\section{The LATIN-PGD solver}\label{sec:LATIN_sol}

Let us consider a concrete medium as illustrated on Figure \ref{fig:ref_prob}.
The mechanical states of the structure are described by different fields $\SOL = \Prth{\vect{u},\ctens{\eps},\ctens{\sigma},d,Y}$ building the solution of the reference problem \ref{PROB:SF}.

From the perspective of writing a weak formulation of the reference problem \ref{PROB:SF} and finding a solution to it, the space $\Su$, where the displacement field is sought, is reduced to:
\begin{equation}
  \Su \eq L^2(\DT;\SuS) \eq \ACC{ \vect{v} : \DT \rightarrow \SuS \ \Big| \ \int_\DT \normi{\vect{v}(t)}^2 dt \ < \ \infty }
\end{equation}
where $\SuS$ denotes the space of kinematically admissible functions defined as:
\begin{equation}
\SuS(\DS;\vect{u}^{D}) = \left\lbrace \vect{u} \in \mathcal{H}(\Omega,\mathbb{R}^{d}) \ | \ \vect{u}(\vx)=\vect{u}^{D}, \hspace{3mm} \forall \vx \in \partial_{D} \DS \right\rbrace
\end{equation}
%
%
%
%
where $\mathcal{H}$ is the Sobolev space defined such as:
\begin{equation}
  \mathcal{H} \prth{\DS;\R^d} \eq \ACC{ \vect{u} : \DS \rightarrow \R^d \ \Big| \ \vect{u} \in L^2(\DS) \ , \ \nabla \vect{u} \in L^2(\DS)}
\end{equation}
and $\normi{\square}$ is the associated $L^2$-norm on $\SuS$.
Denoting then $\SuT = L^2(\DT;\R)$, the space $\Su=L^2(\DT;\SuS)$ can be identified to the tensor product space $\SuS \otimes \SuT$ (see e.g. \cite{hackbusch2012tensor}).

\subsection{Reformulation of the reference problem}
	\label{SEC:3:LATIN_PGD:Reformulation}

The reference problem \ref{PROB:SF} is first reformulated introducing two sub-spaces $\EGamma$ and $\EAd$ as follows.

\Par\emph{Constitutive relations}: Let $\EGamma$ denote the manifold containing the fields $\prth{{\ctens{\varepsilon}}, {\ctens{\sigma}}, {d}, {Y}}$ for which the nonlinear part of the constitutive relations is verified and which reads:
\begin{empheq}[box=\fbox]{equation}
   \label{EQ:LATIN:Egamma}
   \EGamma \eq \ACC{ \ \prth{{{\ctens{\varepsilon}},\ctens{\sigma}}, {d}, {Y}} \  \Big| \ \prth{{\ctens{\sigma},{Y}}} = \mathcal{N} \prth{ {\ctens{\varepsilon}},{d}}  \ , \ \mbox{on } \DS\times\DT }
\end{empheq}
where $\mathcal{N}$ is a function associated to the nonlinear part of the constitutive equations.

\Par\emph{Admissibility}: For each one of the fields $\prth{\vect{u},\ctens{\varepsilon},\ctens{\sigma},d,Y}$, one defines spaces on which the solution fields are declared \emph{admissible}.
Thus, we introduce:
\begin{enumerate}[label=$\bullet$]
   \item the space $\Su(\vect{u}_{0},\dot{\vect{u}}_{0},\vect{u}^{D})$ of the \emph{kinematically admissible} displacement fields $\vect{u} \in \Su$ verifying:
\end{enumerate}
\label{EQ:NL:NINCR:LATIN:Eu}
\begin{equation}
\left\{
   \begin{array}{c}
     \vect{u}|_{t=0} \eq \vect{u}_{0} \\
     \dot{\vect{u}}|_{t=0} \eq \dot{\vect{u}}_{0}
   \end{array}\right.
   \quad \mbox{on } \ \DS \qquad \quad \mbox{and} \qquad  \quad
   \vect{u} \eq \vect{u}^{D} \quad \mbox{on } \ \partial_{D}\DS \times \DT
\end{equation}
\begin{enumerate}[label=$\bullet$]
   \item the space $\Eeps(\vect{u}_{0},\dot{\vect{u}}_{0},\vect{u}^{D})$ of the \emph{kinematically admissible} strain fields $\ctens{\varepsilon}$ associated to a displacement field $\vect{u}$ verifying the initial conditions, such as $\ctens{\varepsilon} = \frac{1}{2}\prth{ \nabla \vect{u} + \nabla \vect{u}^T }$ on the whole domain $\DS \times \DT$ and verifying the Dirichlet boundary conditions in a weak sense:
\end{enumerate}
\begin{equation}
\label{EQ:NL:NINCR:LATIN:Eeps}
    \tableq{
    \begin{array}{l}
    \forall \test{\ctens{\sigma}} \in \Esig(0), \\
    \forall \utest \in \Su(0),
    \end{array}
    }{
     \qquad  - \int\limits_{\DS \times \DT} \test{\ctens{\sigma}} : \ctens{\varepsilon}(\vect{u}) \ d\DS dt + \int\limits_{\partial_{D}\DS  \times\DT} \test{\ctens{\sigma}} \dotp \vect{n} \dotp \vect{u}^{D} \ dS dt \eq \int\limits_{\DS \times \DT} \vect{u} \dotp \rho \utestdtdt \ dS dt
    }
\end{equation}
\begin{enumerate}[label=$\bullet$]
   \item the space $\Esig(\vect{f},\vect{f}^{N})$ of the dynamically admissible stress fields $\ctens{\sigma}$ verifying:
\end{enumerate}
\begin{equation}
\label{EQ:NL:NINCR:LATIN:Esig}
    \tableq{
    \begin{array}{l}
    \forall \utest \in \Su(0), \vect{u} \in \Su(\vect{u}_{0},\dot{\vect{u}}_{0},\vect{u}^{D}), \\
    \\
    \\
    \\
    \end{array}
    }{
    - \int\limits_{\DS \times \DT} \ctens{\sigma} : \ctens{\eps} (\utest) \ d\DS dt + \int\limits_{\DS \times \DT} \rho \vect{f} \dotp \utest \ d\DS dt + \int\limits_{\partial_{N}\DS \times \DT} \vect{f}^{N} \dotp \utest \ dS dt \eq \int\limits_{\DS \times \DT} \utest \dotp \rho \ddot{\vect{u}} \ d\DS dt
    }
\end{equation}

Finally, the space $\EAd$ of admissible solution fields $\Prth{\vect{u},\ctens{\eps},\ctens{\sigma},d,Y}$ is defined as:
\begin{empheq}[box=\fbox]{equation}
   \label{EQ:NL:NINCR:LATIN:EAd}
    \EAd \eq \ACC{ \prth{\vect{u},\ctens{\sigma},\ctens{\eps},d,Y} \quad \Big| \quad \vect{u}\in\Su(\vect{u}_{0},\dot{\vect{u}}_{0},\vect{u}^{D}) \ , \ \ctens{\eps} \in \Eeps(\vect{u}_{0},\dot{\vect{u}}_{0},\vect{u}^{D}) \ , \ \ctens{\sigma} \in \Esig(\vect{f},\vect{f}^{N}) \ \ \mbox{and} \ \ \prth{\ctens{\varepsilon},d} = \mathcal{L}\prth{\ctens{\sigma},Y}}
\end{empheq}
where the function $\mathcal{L}$ is associated to the linear part of the constitutive relations.

\begin{figure}[!ht] 
\centering
\begin{tikzpicture}[line cap=round,line join=round,>=stealth,x=0.9cm,y=0.9cm,scale=1]
    \coordinate (O) at (-0.3,0) ;
    \coordinate (A) at (7,0) ;
    \coordinate (B) at (-0.3,5) ;
    \draw [->][black][line width=1pt][cap=round] (O) -- (A);
    \draw [->][black][line width=1pt][cap=round] (O) -- (B);
    \draw (A) node[below left] {\color{black}{$\ctens{\varepsilon},d$}} ;
    \draw (B) node[below left] {\color{black}{$\ctens{\sigma},Y$}} ;
    \newcommand{\xS}{0.666}
    \draw [domain=0.2:6][black][line width=1.2pt] plot(\x,{2+2*ln(\xS) + 0.37*(\x-\xS)}) ;
    \draw (6,{2+2*ln(\xS) + 0.37*(6-\xS)}) node[below right] {\color{black}$\EAd$} ;
    \draw [domain=0.5:5][black][line width=1.2pt] plot(\x,{2+2*ln(\x)}) ;
    \draw (5,{2+2*ln(5)}) node[right] {\color{black}$\EGamma$} ;
    \node (S) at (\xS,{2+2*ln(\xS)}) {$\scriptstyle\blacksquare$} ;
    \draw (S) node[below right] {\color{black}$\SOLex$} ;
    \newcommand{\xSA}{4.5}
    \newcommand{\xSB}{3}
    \newcommand{\xSC}{2.5}

    \node (SA) at (\xSA,{2+2*ln(\xS) + 0.37*(\xSA-\xS)}) {\color{black}$\bullet$} ;
    \draw (SA) node[below right] {\color{black}{$\SOLa$}} ;

    \node (SB) at (\xSB,{2+2*ln(\xSB)}) {\color{black}$\bullet$} ;
    \draw (SB) node[above left] {\color{black}{$\SOLb$}} ;

    \node (SC) at (\xSC,{2+2*ln(\xS) + 0.37*(\xSC-\xS)}) {\color{black}$\bullet$} ;
    \draw (SC) node[below right] {\color{black}{$\SOLc$}} ;

    \draw [->][black][line width=1.2][cap=round] (SA) -- (SB);
    \draw [->][black][line width=1.2][cap=round] (SB) -- (SC);

    \coordinate (SAB) at ({0.5*(\xSA+\xSB)},{0.5*(2+2*ln(\xS) + 0.37*(\xSA-\xS)+2+2*ln(\xSB))}) ;
    \draw (SAB) node[above right] {\color{black}{$\SDAG$}} ;

 \coordinate (SBC) at ({0.5*(\xSB+\xSC)},{0.5*(2+2*ln(\xSB) + 2+2*ln(\xS) + 0.37*(\xSC-\xS))}) ;

\draw (SBC) node[left] {\color{black}{$\SDGA$}} ;

    \coordinate (SBC) at ({0.5*(\xSB+\xSC)},{0.5*(2+2*ln(\xSB) + 2+2*ln(\xS) + 0.37*(\xSC-\xS))}) ;
\end{tikzpicture}
\caption{Iterative strategy with search directions $\SDaC$ and $\SDdC$.}
\label{FIG:NL:NINCR:LATIN:ITER}
\end{figure}
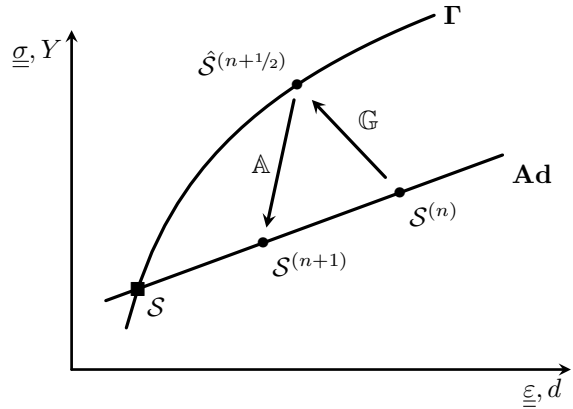

The sought solution $\SOLex$ is then defined at the intersection of the two spaces $\EGamma$ and $\EAd$ (see Fig. \ref{FIG:NL:NINCR:LATIN:ITER} for illustration).


	\subsection{The LATIN method: main ingredients}
	\label{SEC:3:LATIN_PGD:Description}
	
The LATIN methodology then consists in finding the solution $\SOLex \in \EGamma \cap \EAd$ in an iterative process composed of \emph{nonlinear stages} (eventually local in time and space) providing a prediction $\SOLb \in \EGamma$ and \emph{linear stages} (global over the whole space-time domain) providing an approximation $\SOLc \in \EAd$.
The iterative search, starting from the elastic solution $\SOLinit$, can be summarized writing:
\begin{equation}
\SOLinit \in \EAd \ \dots \ \longrightarrow \ \SOLa \in \EAd \ \longrightarrow \ \SOLb \in \EGamma \ \longrightarrow \ \SOLc \in \EAd  \  \dots \ \longrightarrow \SOLex \in \EAd \cap \EGamma
\end{equation}

The jump from one subspace to the other is done by introducing search directions $\SDAG$ and $\SDGA$ as illustrated on Figure \ref{FIG:NL:NINCR:LATIN:ITER}. 
Those search directions are parameters of the LATIN method.
Both linear and nonlinear steps are briefly described in the following lines and explained in details in Section \ref{sec:Implementation}.
More general details on the methodology can be found in \cite{ladeveze1999nonlinear}.

\Par\emph{{Local Stage.}} 
Knowing $\SOLa\in\EAd$, the problem at the local stage consists in finding $\SOLb\in\Egamma$ such that the search direction $\SDAG$ is satisfied.
In the context of our study on reinforced-concrete, this search direction is chosen as:
\begin{equation}
   \label{EQ:NL:NINCR:LATIN:SDAG}
   \SDAG\quad \left\{
   \begin{array}{r}
   \IVYh^{\nb} - \IVY^{\na} \eq 0 \\
   \VVepsh^{\nb} - \VVeps^{\na} \eq 0
   \end{array}\right.
\end{equation}
The choice of the search direction can vary according to the nature of the problem to be solved (see e.g. \cite{ladeveze1999nonlinear,ladeveze2010latin,relun2013model}).
It can be chosen constant or not over the different iterations $(n)$.
In the present case, the local stage boils down to an actualization of the damage pattern $\IVd$ and associated stress field $\VVsig$ over $\DS\times\DT$ according to the nonlinear constitutive relations (see section \ref{SEC:4:Implementation:Local_Step} for details).

\Par\emph{Linear Stage.}
From the knowledge of $\SOLb\in\EGamma$, one now seeks the iterate $\SOLc\in\EAd$ verifying the admissibility conditions and satisfying a new search direction $\SDGA$.
In this work, we choose this direction such as:
\begin{equation}
   \label{EQ:NL:NINCR:LATIN:SDGA}
   \SDGA\quad \left\{
   \begin{array}{r}
   \IVd^{\nc} - \IVdh^{\nb} \eq 0 \\
   \Prth{\VVsig^{\nc} - \VVsigh^{\nb}} \ - \ \SDGA^{\na} \pbic \Prth{\VVeps^{\nc} - \VVepsh^{\nb}} \eq 0
   \end{array}\right.
\end{equation}
Different choices may be investigated for the choice of $\SDGA^{\na}$, for example based on the tangent operator \cite{relun2013model} writing $\SDGA^{\na} \equiv \frac{\partial \mathcal{N}}{\partial \VVeps}|_{\VVeps=\VVepsh}$.
One can note that in the latter case, the operator $\SDGA^{\na}$ must be recomputed for each iteration $\na$, which may lead to some computational over-costs, but in the hope of faster convergence.
In this work, echoing to a quasi-Newton approach (see also \cite{ladeveze2010latin,vitse2019dealing}, a constant identical search direction is chosen using the Hooke tensor $\HookeC$ associated to the undamaged concrete initial configuration, writing simply: $\SDGA^{\na} = \HookeC$.
Analogously to the local stage, we choose the search direction to be constant over the iterations.

As was already explained in the introduction, this stage consists in reapplying dynamic equilibrium and kinematic relations on the whole space-time domain $\DS\otimes\DT$.
The problem to be solved at this stage is then said \emph{global} (in space and time) but is linear.
This linear problem to solve is detailed in sub-section \ref{sec:global_stage_iso}, where we will particularly explain how the solution $\SOL^{\nc}$ can be approximated by a low rank combination of PGD modes, bringing the proposed methodology its numerical efficiency.
Let on note that if the time-domain is of consequent size, computing the temporal functions of the PGD approximation can still be expensive.
In this work a robust and numerically efficient Time Discontinuous Galerkin Method (TDGM) is carried out for the computation of the temporal contribution; details are given in sub-section \ref{sec:global_stage_iso}.


\subsection{Convergence and stop criterion}
	\label{SEC:3:LATIN_PGD:Stop_Criterion}

The convergence of the LATIN algorithm has been proved in quasistatics \cite{ladeveze1999nonlinear} for the case where the mapping $\mathcal{N}$ is monotonic.
The solution $\SOL$ being the intersection of the two spaces $\EAd$ and $\EGamma$, the measure of the distance between two iterates $\SOLb$ and $\SOLc$ is a good error indicator for estimating the convergence of the strategy.
In this study, the following ratio is computed:
\begin{equation}\label{eq:ErrorL_dam}
\LATINerror^{\nc} =  \sqrt{ \left( \frac{ \normST{\ctens{\sigma}^{\nc} - \hat{\ctens{\sigma}}^{\nb}}^2 }{ \normST{\ctens{\sigma}^{\nc}}^2 } + \frac{ \normST{\ctens{\varepsilon}^{\nc} -\hat{\ctens{\varepsilon}}^{\nb}}^2 }{ \normST{\ctens{\varepsilon}^{\nc}}^2 } \right) }
\end{equation}
where the norm $\normST{\bullet}$ reads:
\begin{equation*}
\normST{{\bullet}}^2 = \int_{\DS \times \DT} ({\bullet}) : ({\bullet}) \ d\DS dt
\end{equation*}
Computations are stopped when the above indicator is less than a given threshold.

In order to ensure the convergence of the algorithm for a wider class of constitutive relations, it is recommended in \cite{ladeveze2010latin} to modify the linear step using a so-called relaxation step.
We rename $\SOLrelax\equiv\SOLc$ the solution computed until now and we redefine the current iterate $\SOLc$ by the relation:
\begin{equation}
   \SOLc  \eq \mu \SOLrelax + (1-\mu) \SOLa
\end{equation}
where the currently used value for the parameter $\mu$ is $0.4$.
This parameter is problem-dependent and could be different for other kind of applications;  finding a correct value would require empirical numerical tests in such case.

\section{Implementation details}\label{sec:Implementation}

The solution of the reformulated reference problem is now approximated using classical finite elements in space and time-discontinuous Galerkin method in time.
The domain $\DS$ is discretized using $\NS$ finite elements, involving $\nsg$ spatial Gauss points whereas the temporal domain is discretized using $\NT$ discontinuous finite elements involving $\ntg$ Gauss points and cubic Hermite polynomials as temporal shape functions.
The order three is chosen for the temporal interpolation of the displacement field for having a linear approximation of the acceleration field. Technical details are not given in this paper, the interested reader is referred to more classical references on finite element integration such as \cite{zienkiewicz2000finite} or \cite{hughes2012finite}.

\subsection{Local Stage}
\label{SEC:4:Implementation:Local_Step}

Assuming that the finite element problem has been discretized in space and time, from the knowledge of the global solution $\SOLa$, the mechanical fields of the new iterate on $\EGamma$ must be computed at all spatial and temporal Gauss points. 
Regarding the material nonlinearity considered in this paper and presented section \ref{sec:iso_damage_cr}, computations performed during the local stage are summarized on Algorithm \ref{ALG:LATIN:Local}.
Let us note, from an implementation point of view, that the constitutive relations can be solved independently for each spatial Gauss points which enables, for this local step, a high degree of parallelization.

\SetKwInput{Entrees}{Inputs}
\SetKwInput{Sorties}{Outputs}
\SetKwInOut{Parametre}{Parameter}

\RestyleAlgo{boxed}
\SetAlCapSkip{1em}
\begin{algorithm}[!ht]
\LinesNumbered
\DontPrintSemicolon
\BlankLine
\Entrees{$\prth{\VVeps,\VVsig,\IVd,\IVY,\IVz,\IVZ}$ \quad iteration $\na$}
\Sorties{$\prth{\VVepsh,\VVsigh,\IVdh,\IVYh,\IVzh,\IVZh}$ \quad iteration $\nb$}
\Parametre{Search direction $\SDAG$}
\BlankLine
\hrule
\BlankLine
Loop over spatial Gauss points\;
\For{$\gS = 1:\NgS$}{
   Loop over temporal Gauss points \;
   \For{$\gT = 1:\NgT$}{
   From the search direction $\SDAG$ \eqref{EQ:NL:NINCR:LATIN:SDAG}, compute the released energy in concrete medium \;
   $\VVepsh^{\nb} = \VVeps^{\na}$  \;
   $\IVYh^{\nb} = \IVY^{\na} = \frac{1}{2}\mbraP{\VVeps^{\na}}\pbic\HookeC\pbic\mbraP{\VVeps^{\na}}$ \;
   Computation of the damage threshold \;
   $\fDth_c^{\nb} = \IVY^{\na} - \prth{\MPYs + \IVZ^{\na}}$ \;
   \eIf{$\fDth_c^{\nb} > 0$}{
   Actualization of damage variables \;
   $\bar{\IVd}^{\nb} \eq 1 - \frac{1}{1+\MPAd\prth{\IVYh^{\nb} - \MPYs}}$ \;
   $\IVzh^{\nb} = -\bar{\IVd}^{\nb}$ \;
   $\IVZh^{\nb} = \frac{1}{\MPAd}\Prth{-1 + \frac{1}{1+\IVz^{\nb}}}$ \;
   }{
   Damage remains constant \;
   $\bar{\IVd}^{\nb} = \bar{\IVd}^{\na}$ \;
   $\IVzh^{\nb} = \IVz^{\na}$ \;
   $\IVZh^{\nb} = \IVZ^{\na}$
   }
   Regularization using damage delay technique, solving \;
   $\dIVd \eq \frac{1}{\MPtauC} \PRTH{1 - \exp\Prth{-\MPaC\prth{\MbraP{\bar{\IVd}^{\nb} - \IVd}}}} \quad \longrightarrow \quad \IVdh^{\nb}$ \;
   Computation of stress in concrete medium + crack -closure part \eqref{eq:closure_relation_eps} \;
   $\VVsigCh^{\nb} = (1-\IVdh^{\nb}) \ \HookeC\pbic\VVepsh^{\nb} \ + \ \IVdh^{\nb} \ \HookeC\pbic \Fcc(\VVepsh^{\nb})$ \;
}
}
\caption{Local stage of the LATIN solver -- Nonlinear part of the concrete constitutive relations.}\label{ALG:LATIN:Local}
\end{algorithm}

\subsection{Global linear stage: equilibrium and compatibility equations}\label{sec:global_stage_iso}

As the solver progresses, for each global stage, we are led to solve a sequence of linear problems written in the space-time domain. 
The problem being linear, a low-rank approximation of the solutions can be efficiently computed using the Proper Generalized Decomposition (PGD) where spatial and temporal functions need to be computed. 
Here, the determination of the spatial functions follows a classic FE approach, but the temporal functions are determined using a less classical Time-Discontinuous Galerkin Method (TDGM). 
This new approach allows to decrease the size of the discretized operators needed to be inverted for the determination of the temporal functions, decreasing therefore computational expenses. 
The following lines detail the different steps required for enriching the decomposition, approaching the solution of the global stage.

The problem to be solved at each global stage consists in finding $\SOLc \in \EAd$ given the knowledge of $\SOLb$. 
As mentioned before, this solution must verify the admissibility conditions of the reference problem as well as the descent search direction \eqref{EQ:NL:NINCR:LATIN:SDGA}, which can be summarized writing:
\begin{empheq}[box=\fbox]{equation}\label{eq:d_sd_damage}
\text{Find} \quad \SOLc \in \EAd \quad \text{such that} \quad \left( \SOLc -  \SOLb \right) \in \SDGA
\end{empheq}

The current solution at iteration $\nc$ that is sought in $\EAd$ is kinematically admissible to zero and must verify the following global equilibrium relation:

\eqtext{$\forall \utest \in \SuS(\DS,0) \otimes \SuT(I) \com$}
\begin{equation}\label{eq:eq_gs}
\int_{\DS \times I} \rhoc\ddot{\vect{u}}^{\nc} \dotp \utest \ d\DS dt +
\int_{\DS \times I}\ctens{\sigma}^{\nc} : \ctens{\varepsilon}(\utest)\hspace{1mm} d\DS dt = \int_{\DS \times I} \rho \vect{f} \dotp \utest \ d\DS dt + \int_{\partial_{N}\DS \times I} \vect{f}^{N} \dotp \utest \ dSdt
\end{equation}

Considering the last computed solution iterate $\SOLa$, one can introduce the current corrective term $\rmDelta \mathcal{S}^{\nc}$ such that:
\begin{equation}\label{eq:corr_terms_iso}
\rmDelta \mathcal{S}^{\nc} \deq \SOLc - \SOLa \qquad \longrightarrow \qquad \ACC{ \rmDelta \vect{u}^{\nc}, \rmDelta \ctens{\varepsilon}^{\nc}, \rmDelta \ctens{\sigma}^{\nc} , \rmDelta\IVd^{\nc}, \rmDelta\IVY^{\nc} }
\end{equation}
containing all the corrective solution fields of the mechanical problem.

The global problem can then be rewritten using the corrective term that is sought in the admissibility space $\EAd$ thus satisfying the equilibrium equation:

\eqtext{$\forall \utest \in \SuS(\DS,0) \otimes \SuT(I) \com$}
\begin{equation}\label{EQ:Global:Equilibrium:dS}
\int_{\DS \times I} \rhoc \rmDelta\ddot{\vect{u}}^{\nc} \dotp \utest \ d\DS dt + \int_{\DS \times I} \rmDelta\ctens{\sigma}^{\nc} : \ctens{\varepsilon}(\utest) \ d\DS dt \eq 0
\end{equation}
and verifying the search direction as closely as possible:
\begin{equation}
\label{EQ:Global:SD:dS}
\rmDelta \ctens{\sigma}^{\nc} - \HookeC : \rmDelta \ctens{\varepsilon}^{\nc} + \ctens{\rmDelta}^{\nc} \eq 0
\end{equation}
The search direction then takes the form of an affine constitutive relation where the loading term simply writes:
\begin{equation}\label{eq:Delta}
\ctens{\rmDelta}^{\nc} \eq \Prth{\ctens{\sigma}^{\na} - \hat{\ctens{\sigma}}^{\nb}}
\end{equation}
Let also recall that $\IVd^{\nc} = \IVdh^{\nb}$, according to the search direction \eqref{EQ:NL:NINCR:LATIN:SDGA} chosen in this work.

The global stage can the be re-written using $\rmDelta \SOLc$ as a minimization problem under constraint where one seeks \emph{dynamically} and \emph{kinematically admissible} space-time corrections $\rmDelta\VVsig^{\nc}$ and $\rmDelta\VVeps^{\nc}$ such that:
\begin{equation}
\label{EQ:Global:MinimizationPB}
\Acc{\rmDelta\VVsig^{\nc},\rmDelta\VVeps^{\nc}} \eq 
\arg \min_{
\begin{array}{c}
\rmDelta\VVsig \in \mbox{DA} \\ \rmDelta\VVeps \in \mbox{KA}
\end{array}
}
\underbrace{
\NormST{\rmDelta \ctens{\sigma} - \HookeC : \rmDelta \ctens{\varepsilon} + \ctens{\rmDelta}^{\nc}}^2_{\HookeC^{-1}}
}_{\displaystyle \CREc}
\end{equation}
where the space-time norm involved for measuring the residual to minimize is defined as:
\begin{equation}
\label{EQ:Global:STnormE}
{\big|\big|\big|{ \ \bullet \ }\big|\big|\big|}_{\HookeC^{-1}}^{2} \eq  \int_{\DS \times \DT} (\bullet): \HookeC^{-1} : (\bullet) \ d\DS dt
\end{equation}
Expression \eqref{EQ:Global:MinimizationPB} is denoted constitutive relation error (CRE) \cite{ladeveze1989methode}.

In this work, in accordance with the classical LATIN-PGD framework \cite{ladeveze1989methode}, the solution fields of the successive iterates $\SOLc$ and yielding corrections $\rmDelta \SOLc$ minimizing problem \eqref{EQ:Global:MinimizationPB} are sought as a low rank approximation using the \emph{Proper Generalized Decomposition} (PGD).
In comparison to what can be done in low-frequency dynamics using classical step-by-step integration where the dynamic equilibrium is projected on a truncated modal basis, in the proposed approach, the truncated PGD-basis is successively enriched to adapt to the increasing nonlinearities involved along the LATIN iterations (see \cite{Daby2022} for a more complete discussion on modal and PGD space basis).
The nonlinear corrections to the elastic solution $\SOLinit$ are then sought as a combination of $\PGDm$ space-time modes such that the solution-fields at iteration $\na$ can be written as:
\begin{equation}
\label{EQ:decomposition}
\begin{split}
\vect{u}^{\na}(\vx,t) &\approx \sum_{i=1}^{\PGDm} \pgdu_{i}(\vx) \pgdt_{i}(t) + \uo(\vx,t)
\\
\ctens{\varepsilon}^{\na}(\vx,t) &\approx \sum_{i=1}^{\PGDm} \pgdeps_{i}(\vx) \pgdt_{i}(t)  + \epso(\vx,t)
\\
\ctens{\sigma}^{\na}(\vx,t) &\approx \sum_{i=1}^{\PGDm} \pgdsig_{i}(\vx) \CharboModif{mycolor}{\mu_{i}(t)} + \sigmao(\vx,t)
\end{split}
\end{equation}

where $\uo$, $\epso$ and $\sigmao$ respectively denote the displacement-, strain- and stress-fields of the kinematically admissible elastic solution $\SOLinit$.
The corrections to this elastic initial solution are sought as a kinematically admissible to zero low rank approximation using the PGD.
Because stress and strain fields are not linked by linear relations, please note that different temporal functions $\prth{\lambda_i}_{i=1}^m$ and $\prth{\mu_i}_{i=1}^m$ are introduced in the decomposition \eqref{EQ:decomposition}.
In contrast, assuming small perturbations, displacement and strain are linearly related and can be described with the same set of time-functions $\prth{\lambda_i}_{i=1}^m$.
\color{black}

Here the PGD decomposition is calculated in a so-called \emph{Enrichment step}. This step consists of determining a new spatio-temporal mode and adding it to the existing PGD decomposition. 

Classically and additional step called \emph{Updating step} is also performed \cite{passieux2008approximation,nachar2019multi}. This step consists in updating the temporal functions only ($\prth{\pgdt_i}_{i=1}^m$ and $\prth{\mu_i}_{i=1}^m$) from the knowledge of the current truncated spatial basis $\prth{\pgdu_i}_{i=1}^m$, $\prth{\pgdeps_i}_{i=1}^m$, $\prth{\pgdsig_i}_{i=1}^m$. However, this technique is not considered in this work since its implementation in the present problem did not significantly improve the algorithm. Further work to improve this step is considered.

The following sub-section detail the enrichment step of this global stage.

\subsubsection{Enrichment step: addition of a new rank-one space-time PGD mode to the existing decomposition}\label{sec:Enrch_concrete}


The enrichment step consists in the calculation of a new PGD mode, of index $\PGDm+1$, for describing each global quantity. 
Those rank-one corrections from solution $\SOLa$ to $\SOLc$ are defined such as:
\begin{equation}
\label{eq:corrective_terms_PGD}
\begin{split}
\rmDelta \vect{u}^{\nc}(\vx,t) &\eq \vect{u}^{\nc}-\vect{u}^{\na} \eq \pgdu_{\PGDm+1}(\vx) \ \pgdt_{\PGDm+1}(t)\\
\rmDelta \ctens{\varepsilon}^{\nc}(\vx,t) &\eq \ctens{\varepsilon}^{\nc} - \ctens{\varepsilon}^{\na} \eq \pgdeps_{\PGDm+1}(\vx) \ \pgdt_{\PGDm+1}(t)
\\
\rmDelta \ctens{\sigma}^{\nc}(\vx,t) &\eq \ctens{\sigma}^{\nc} - \ctens{\sigma}^{\na} \eq \pgdsig_{\PGDm+1}(\vx) \ \CharboModif{mycolor}{\mu_{\PGDm+1}(t)}
\end{split}
\end{equation}
\color{mycolor}
and the reference minimization problem \eqref{EQ:Global:MinimizationPB} becomes:
\begin{equation}
\label{EQ:Global:MinimizationPB:Enrich}
\Acc{\pgdsig_{\PGDm+1}\mu_{\PGDm+1} \ , \ \pgdeps_{\PGDm+1}\pgdt_{\PGDm+1}} \eq 
\arg \min_{
\begin{array}{c}
\pgdsig\mu \in \mbox{DA} \\ \pgdeps\pgdt \in \mbox{KA}
\end{array}
}
\underbrace{
\NormST{\pgdsig\mu - \HookeC : \pgdeps\pgdt + \ctens{\rmDelta}^{\nc}}^2_{\HookeC^{-1}}
}_{\displaystyle \CREc}
\end{equation}
\color{black}

The new spatial $\overline{\square}_{\PGDm+1}$ and temporal $(\pgdt_{\PGDm+1},\mu_{\PGDm+1})$ contributions are computed using a fixed point algorithm with alternate search directions as explained in the following lines.
The temporal functions $(\pgdt_{\PGDm+1},\mu_{\PGDm+1})$ are randomly initialized and the spatial functions $\overline{\square}_{\PGDm+1}$ are computed from the first initial values of $(\pgdt_{\PGDm+1},\mu_{\PGDm+1})$.
From the knowledge of $\overline{\square}_{\PGDm+1}$, the contributions $(\pgdt_{\PGDm+1},\mu_{\PGDm+1})$ are then computed over the whole temporal domain, and so on.
A cascade sequence of spatial and temporal problems is solved until stagnation of the corrections.

For checking this stagnation, spatial functions are first normalized with respect to the norm of the deformation tensor.
One then defines:
\begin{equation}
c_{c} = \normi{\pgdeps_{\PGDm+1}}_{\DS} \quad \mbox{with} \quad \normi{\bullet}_{\DS} = \sqrt{ \int_{\DS} (\bullet):(\bullet) \ d\DS  }
\end{equation}
and PGD space- and time-functions are rewritten such as:
\begin{equation*}
\begin{split}
\pgdu_{\PGDm+1} &\leftarrow \pgdu_{\PGDm+1} / c_{c} \\
\pgdeps_{\PGDm+1} &\leftarrow \pgdeps_{\PGDm+1} / c_{c}\\
\pgdsig_{\PGDm+1} &\leftarrow \pgdsig_{\PGDm+1} / c_{c}
\end{split}
\quad \quad , \quad \quad
\begin{split}
\pgdt_{\PGDm+1} &\leftarrow \pgdt_{\PGDm+1} \ c_{c} \\
\dot{\pgdt}_{\PGDm+1} &\leftarrow \dot{\pgdt}_{\PGDm+1} \ c_{c} \\
\ddot{\pgdt}_{\PGDm+1} &\leftarrow \ddot{\pgdt}_{\PGDm+1} \ c_{c}
\end{split}
\CharboModif{mycolor}{
\quad \quad , \quad \quad
\begin{split}
\mu_{\PGDm+1} &\leftarrow \mu_{\PGDm+1} \ c_{c}
\end{split}
}
\end{equation*}
The stagnation error of the fixed-point algorithm is defined as:
\begin{equation}
\pgderror = \frac{ \normi{  \abs{\pgdt_{\PGDm+1}^{(i)}} - \abs{\pgdt_{\PGDm+1}^{(i-1)}}    }_{\DT}}{ \normi{ \abs{\pgdt_{\PGDm+1}^{(i)}} + \abs{\pgdt_{\PGDm+1}^{(i-1)}}   }_{\DT}} \quad \mbox{with} \quad
\normi{\bullet}_{\DT} = \sqrt{ \int_{\DT} (\bullet)^2 dt }
\end{equation}

The functions $\pgdt_{\PGDm+1}^{(i)}$ and $\pgdt_{\PGDm+1}^{(i-1)}$ stand for the temporal mode at iteration $(i)$ and $(i-1)$ respectively (integer $(i)$ counts the fixed-point algorithm iterations). 
The fixed point strategy stops when the stagnation indicator reaches a given threshold. 
At the end of the process the spatial and temporal functions are kept and added to the current low-rank approximation.

The solution fields $\Acc{\pgdsig_{\PGDm+1}\mu_{\PGDm+1},\pgdeps_{\PGDm+1}\pgdt_{\PGDm+1}}$ are sought as the solution of a minimization problem under constraints \eqref{EQ:Global:MinimizationPB:Enrich} which could be done using Lagrange multipliers.
However in this work, for computational efficiency, the space functions of the decomposition are determined by a Galerkin projection of the equilibrium equation \eqref{EQ:Global:Equilibrium:dS} integrating the search direction, while the temporal contributions are determined using a Galerkin projection on the equilibrium equation for $\pgdt(t)$ and by minimizing the constitutive relation error for $\mu(t)$. 
These steps are detailed in the following lines, where, the PGD index $(\PGDm+1)$ is omitted to simply write:
\begin{equation}\label{eq:sm_c_ad}
\begin{split}
\rmDelta \vect{u}^{\nc}(\vx,t)&\eq  \pgdu(\vx) \pgdt(t) \\
\rmDelta \ctens{\varepsilon}^{\nc}(\vx,t) &\eq \pgdeps(\vx) \pgdt(t) \\
\rmDelta \ctens{\sigma}^{\nc}(\vx,t) &\eq \pgdsig(\vx) \CharboModif{mycolor}{\mu(t)}
\end{split}
\end{equation}

\noindent $\bullet$ \textbf{Space PGD functions determination}:\\

From the knowledge of the temporal functions $\pgdt$ and $\mu$, we aim at computing the space functions $\pgdu$, $\pgdeps$ and $\pgdsig$.
As was already explained, a simplified solution of the minimization problem \eqref{EQ:Global:MinimizationPB} is sought.
The residual on the search direction $\SDGA$ in \eqref{EQ:Global:MinimizationPB} must be minimized; here we consider it to be verified and introduce the search direction \eqref{EQ:Global:SD:dS} directly into the dynamic equilibrium \eqref{EQ:Global:Equilibrium:dS} to yield:

\eqtext{$\forall \utest \in \SuS(\DS,0) \otimes \SuT(I) \com$}
\begin{equation}\label{EQ:Global:Enrichment:EquilibriumSD}
\int_{\DS \times I} \rhoc \rmDelta\ddot{\vect{u}}^{\nc} \dotp \utest \ d\DS dt + \int_{\DS \times I} \ctens{\varepsilon}(\utest) : \HookeC : \rmDelta \ctens{\varepsilon}^{\nc} \ d\DS dt \eq \int_{\DS \times I} \Lres^{\nc} : \ctens{\varepsilon}(\utest)\hspace{1mm} d\DS dt
\end{equation}
Introducing then the low-rank approximation of the correction \eqref{eq:sm_c_ad} can then be introduced into the equilibrium equation \eqref{EQ:Global:Enrichment:EquilibriumSD} we obtain:

\eqtext{$\forall \pgdv \in \SuS(\DS,0) \com$}
\begin{equation}\label{eq:space_pgd}
\int_{\DS} \rhoc \pscT{\ddot{\pgdt}\pgdt} \pgdu \dotp \pgdv \hspace{1mm}d\DS + \int_{\DS} \pscT{\pgdt\pgdt} \pgdeps(\pgdv) : \HookeC : \pgdeps(\pgdu) \ d\DS  = \int_{\DS} \pscT{\Lres^{\nc}\pgdt} : \pgdeps(\pgdv) \hspace{1mm} d\DS
\end{equation}
where we denote $\pscT{\bullet} = \int_{\DT} (\bullet) \ dt$ to simplify the notations.

The above weak formulation can be discretized using classical FEM and the unknown displacement field $\pgdu$ can be computed after simple matrix inversion. 
From the knowledge of the corrective spatial displacement field $\pgdu(\vx) \in \SuS(\DS,0)$, the spatial corrective strain $\pgdeps(\pgdu)$ can be reconstructed.

The spatial function associated to the corrective stress $\pgdsig$ must now be calculated and is determined by minimizing \eqref{EQ:Global:SD:dS}, resulting in:
\begin{equation}
\pgdsig(\vx) =  \frac{ \HookeC : \pgdeps(\vx) \pscT{\mu \pgdt} - \pscT{\mu \Lres^{\nc}} }{ \pscT{\mu^2} }
\end{equation}

\color{black}

Once the space functions $\pgdu(\vx)$, $\pgdeps(\vx)$ and $\pgdsig(\vx)$ are computed, the temporal functions $\pgdt(t)$ and $\mu(t)$ must be determined. 
Associated developments are shown below.

\Par

\noindent $\bullet$ \textbf{Temporal determination of PGD functions using TDGM}:\\

From the knowledge of the space functions $\pgdu$, $\pgdeps$ and $\pgdsig$, we now aim at computing the temporal functions $\pgdt(t)$ and $\mu(t)$. At one hand, the temporal function associated to the displacement and strain ($\pgdt(t)$) is determined by verifying the equilibrium equation, this is, by considering \eqref{EQ:Global:Enrichment:EquilibriumSD} and integrating in space one obtains:
\begin{equation}\label{eq:temp_eq}
a \int_{\DT} \pgdt^* \ddot{\pgdt} dt + b \int_{\DT} \pgdt^* \pgdt dt = \int_{\DT} \pgdt^* f(t) dt 
\end{equation}
where one has the scalars $a$ and $b$, and the temporal function $f(t)$ expressed as follows:
\begin{align}
a &= \int_{\DS} \rhoc \pgdu \dotp \pgdu \hspace{1mm} d\DS \\
b &= \int_{\DS} \pgdeps(\pgdu) : \HookeC : \pgdeps(\pgdu) d\DS \\
f(t) &= \int_{\DS} \Lres^{\nc} : \pgdeps(\pgdu) \hspace{1mm} d\DS
\end{align}
On the other hand, since the corrective term associated to the stress tensor should minimize the CRE of \eqref{EQ:Global:MinimizationPB}, the time function $\mu(t)$ is determined by solving the following problem:
\begin{equation}\label{eq:min_mu}
\lbrace  \mu \rbrace = \underset{\mu \in \SuT}{\text{arg min}} \norm{ \pgdsig \mu - \HookeC : \pgdeps \pgdt  + \Lres^{\nc} }_{\HookeC^{-1}}^{2}
\end{equation}
Here, in order to solve \eqref{eq:temp_eq} and \eqref{eq:min_mu} at low cost, the TDGM is used. The idea is to solve incrementally the function over the temporal domain using at the same time a finite element formulation. To illustrate this idea, one considers below the TDGM for the determination of $\pgdt(t)$ only, the procedure applied to $\mu(t)$ is similar and is not considered here.

The temporal FEM discretization of expression \eqref{eq:temp_eq} applied at a time element ``$k$" is given as:
\begin{equation}
\label{eq:discr_ori}
\Mt^{[k]} \p \SDvect{\pgdt}^{[k]} = \Ft^{[k]}
\end{equation}
with:
\begin{equation}\label{Time_min_terms_discr}
\Mt^{[k]} = \int_{\breve{I}_{k}} a \ \sfTv^{[k]}(t) \otimes \ddot{\sfTv}^{[k]}(t) + b \ \sfTv^{[k]}(t) \otimes \sfTv^{[k]}(t) dt \quad , \quad 
\Ft^{[k]} =\int_{\breve{I}_{k}} \sfTv^{[k]}(t)  f(t) dt 
\end{equation}
and where $\SDvect{\pgdt}^{[k]}$ corresponds to the nodal values of the temporal function at time element $k$. From \eqref{Time_min_terms_discr} one also considered $\sfTv^{[k]}$ the FEM shape functions at element $k$ and $\otimes$ the Kronecker product.

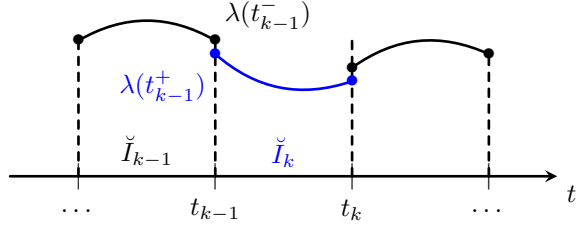
\begin{figure}[H]
\centering
\begin{tikzpicture}[line cap=round,line join=round,>=stealth,x=1.0cm,y=1.0cm,scale=0.9]
  \newcommand{\ysp}{-0.5}
  \draw [->][black][line width=1pt] (-4,0) -- (4,0);
  \draw (4,0) node[below right] {\color{black}$t$} ;
  \draw (0,0) node[above] {\color{blue}$\DTk$} ;
  
  \draw (-2,0) node[above] {\color{black}$\DTd_{k-1}$} ;

  \draw (-3,0) node {\color{black}$|$} ;
  \draw (-3,\ysp) node {\color{black}$\dots$} ;
  \draw (-1,0) node {\color{black}$|$} ;
  \draw (-1,\ysp) node {\color{black}$t_{k-1}$} ;
  \draw (+1,0) node {\color{black}$|$} ;
  \draw (+1,\ysp) node {\color{black}$t_{k}$} ;
  \draw (+3,0) node {\color{black}$|$} ;
  \draw (+3,\ysp) node {\color{black}$\dots$} ;
  \draw [dashed][black][line width=1] (-3,0) -- (-3,2);

  \draw (-3,2) node {\color{black}$\bullet$} ;
  \draw [dashed][black][line width=1] (-1,0) -- (-1,2);
  \draw (-1,2) node {\color{black}$\bullet$} ;
  \draw [-][black][line width=1] (-3,2) to[bend left] (-1,2);


  \draw (-1,1.8) node {\color{blue}$\bullet$} ;
  \draw [dashed][black][line width=1] (+1,0) -- (+1,2);
  \draw (+1,1.4) node {\color{blue}$\bullet$} ;
  \draw [-][blue][line width=1] (-1,1.8) to[bend right] (+1,1.4);


  \draw (+1,1.6) node {\color{black}$\bullet$} ;
  \draw [dashed][black][line width=1] (+3,0) -- (+3,1.8);
  \draw (+3,1.8) node {\color{black}$\bullet$} ;
  \draw [-][black][line width=1] (+1,1.6) to[bend left] (+3,1.8);
  

  \draw (-1,1.7) node[below left] {\color{blue}$\pgdt(t_{k-1}^+)$} ;
  \draw (-1,2.0) node[above right] {\color{black}$\pgdt(t_{k-1}^-)$} ;

\end{tikzpicture}
\caption{Weak imposition of the continuity between the intervals.}
\label{fig:Discr_Time_PGD_enrch}
\end{figure}

However, this equation is incomplete since it does not impose continuity between time intervals. To solve this issue, continuity between time elements is imposed in a weak sense as illustrated in Figure \ref{fig:Discr_Time_PGD_enrch}.
The idea consists in transmitting the end value of the time function of the interval $\DTd_{k-1}$ ($\pgdt(t_{k-1}^{-})$) to the initial value of the interval $\DTd_{k}$ ($\pgdt(t_{k-1}^{+})$), by using some operators that must be included in the discretized equation \eqref{eq:discr_ori} (see Figure \ref{fig:Discr_Time_PGD_enrch} ). The easiest way to do this consists in defining the following operators:
\begin{equation}
\begin{split}
\LM^{[k]} &= 1.1 \hspace{0.2mm}\max( \Mt^{[k]} )
\sfTv^{[k]}(t_{k-1}) \otimes \sfTv^{[k]}(t_{k-1}^{+}) \\
\RM^{[k]} &= 1.1 \hspace{0.2mm}\max( \Mt^{[k]} ) \sfTv^{[k]}(t_{k-1})\otimes \sfTv^{[k-1]}( t_{k-1}^{-})
\end{split}
\end{equation}
whose detailed representations are given by:
\begin{equation*}
\LM^{[k]} = \begin{bmatrix}
1.1\max(\Mt^{[k]}) & 0 & 0 & 0 \\
0 & 0 & 0 & 0 \\
0 & 0 & 0 & 0 \\
0 & 0 & 0 & 0 
\end{bmatrix} \quad , \quad \RM^{[k]} = \begin{bmatrix}
0 & 0 & 0 & 1.1\max(\Mt^{[k]}) \\
0 & 0 & 0 & 0 \\
0 & 0 & 0 & 0 \\
0 & 0 & 0 & 0 
\end{bmatrix}
\end{equation*}
These operators are introduced in the discretized equation of each time interval $\DTd_{k}$ in the following way:
\begin{equation}\label{eq:min_prob_discr_iso}
\left( \Mt^{[k]} + \LM^{[k]} \right) \p  \SDvect{\pgdt}^{[k]} = \RM^{[k]} \p \SDvect{\pgdt}^{[k-1]}  + \Ft^{[k]}
\end{equation}
The factor $1.1\max(\Mt^{[k]})$, with $\max(\cdot)$ the function that finds the maximum value of the matrix $(\cdot)$, is chosen in order to add terms in the same order of magnitude compared to the discretized elemental matrix $\Mt^{[k]}$ of the problem. These operators allows to unbalance the equation  associated to the first nodal value of $\SDvect{\pgdt}^{[k]}$ (first row) of equation \eqref{eq:min_prob_discr_iso},  such as the only way of approximately solving \eqref{eq:discr_ori} is to verify $\LM^{[k]} \p \SDvect{\pgdt}^{[k]} \approx \RM^{[k]} \p  \SDvect{\pgdt}^{[k-1]}$, which is traduced in $\pgdt(t_{k-1}^{-}) \approx \pgdt(t_{k-1}^{+})$. 

Thanks to this formulation, the time functions are solved incrementally, one temporal FEM element at a time. Additionally, the matrix $\Mt^{[k]}$ remains the same $\forall k$ since $a$ and $b$ are scalars (the same happens for the determination of $\mu(t)$), so the inverse of $(\Mt^{[k]} + \LM^{[k]})$ can be computed once and then reused during resolution, thus reducing the computational costs for the determination of the time functions.

\section{Numerical example}
\label{sec:Num_Results}

\noindent 
The capabilities of the LATIN-PGD methodology for solving low-dynamics problems is now illustrated using a numerical example. 
This test-case consists of a 3D bending beam in dynamics subjected to a vertical displacement ($z$ direction) on both sides of the beam and corresponding zero displacements in the other directions as illustrated on Figure \ref{fig:test_case4}.
\def\disto{4}
\def\longaxes{1.5}
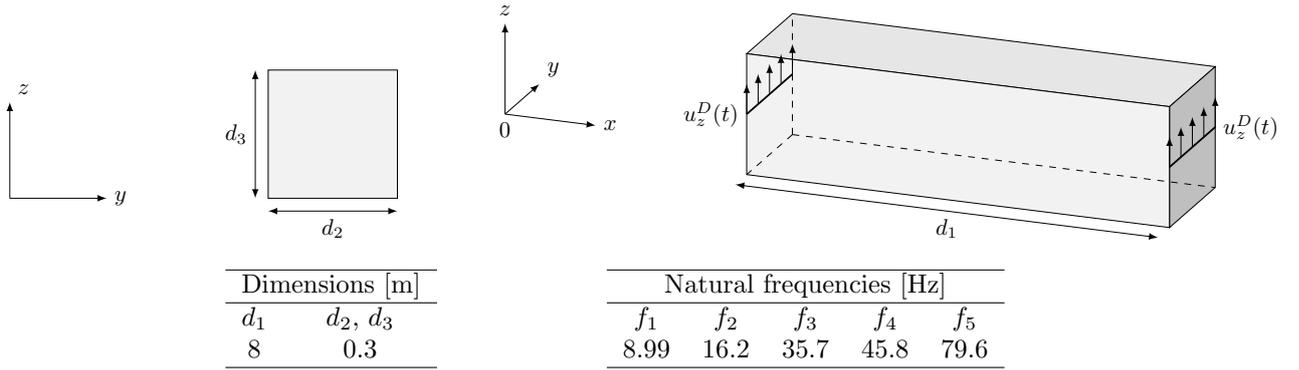
\begin{figure}[H] 
\centering
\begin{center}
\begin{subfigure}{0.35\textwidth}
\scalebox{0.85}{
\begin{tikzpicture}
    \node (origin) at (0,0) {}; 

\def\disto{4}
\def\longaxes{1.5}

 \coordinate (O) at (-\disto,0);

\draw[axis] (O) -- ++(\longaxes,0) node[right] {$y$};
        \draw[axis] (O) -- ++(0,\longaxes) node[above right] {$z$};

    \coordinate (O1) at (0,0);

\def\side{2.0}
\draw[fill=gray!10] (0,0) -- (\side,0) -- (\side,\side) -- (0,\side) -- cycle;

 \def\rmDelta{0.2}
 \draw[dim] (O1) ++ (0,-\rmDelta) -- ++(\side,0) node[midway,below] {$d_2$};
 \draw[dim] (O1) ++(-\rmDelta,0) -- ++(0,\side) node[midway,left] {$d_3$};

  \end{tikzpicture}%
  }
\end{subfigure}
\begin{subfigure}{0.55\textwidth}
\scalebox{0.85}{
\begin{tikzpicture}[dimetric2] %

        \coordinate (O) at (-\disto,0,0);
        \draw[axis] (O) -- ++(\longaxes,0,0) node[right] {$x$};
        \draw[axis] (O) -- ++(0,\longaxes,0) node[above right] {$y$};
        \draw[axis] (O) -- ++(0,0,\longaxes) node[above] {$z$};

\draw (O) node[below] {\color{black}{$0$}} ;







    \def\high{1.5} 

    \def\dx{7}
    \def\dy{2}
    \def\dz{2}

    \draw[fill=gray!20] (0,0,\high) -- (\dx,0,\high) -- (\dx,+\dy,\high) -- (0,+\dy,\high) -- cycle;


    \draw[fill=gray!50] (\dx,0,\high-\dz) -- (\dx,+\dy,\high-\dz) -- (\dx,+\dy,\high) -- (\dx,0,\high) -- cycle;

    \draw[fill=gray!10] (0,0,\high-\dz) -- (\dx,0,\high-\dz) -- (\dx,0,\high) -- (0,0,\high) -- cycle;


\coordinate (O1) at (0,\dy,\high-\dz);

\draw[axis,dashed,-] (O1) -- ++(0,0,\dz);

\draw[axis,dashed,-] (O1) -- ++(0,-\dy,0);

\draw[axis,dashed,-] (O1) -- ++(\dx,0,0);




\foreach \y in {0,0.5,...,\dy} {
      \draw[-latex] (0,\y,\high-\dz*0.5) -- ++(0,0,0.5);
    }

    \foreach \y in {0,0.5,...,\dy} {
      \draw[-latex] (\dx,\y,\high-\dz*0.5) -- ++(0,0,0.5);
    }

\draw[thick] (0,0,\high-\dz*0.5) -- (0,\dy,\high-\dz*0.5); 
\draw[thick] (\dx,0,\high-\dz*0.5) -- (\dx,\dy,\high-\dz*0.5); 

    \draw (0,0,\high-\dz*0.5)  node[above,left] {$u^{D}_{z}(t)$};

    \draw (\dx,\dy,\high-\dz*0.5) node[above,right] {$u^{D}_{z}(t)$};

\draw[dim] (0,-0.5,\high-\dz) -- ++(\dx,0,0) node[midway,below] {$d_1$};





  \end{tikzpicture} %
  }
\end{subfigure}
\end{center}
\begin{tabular}{cc}
\hline
\multicolumn{2}{c}{Dimensions [m]} \\
\hline
$d_1$ & $d_2$, $d_3$ \\
8 & 0.3 \\
\hline
\end{tabular}
\hspace{2cm}
\begin{tabular}{ccccc}
\hline
\multicolumn{5}{c}{Natural frequencies [Hz]} \\
\hline
$f_1$ & $f_2$ & $f_3$ & $f_4$ & $f_5$ \\
{8.99} & {16.2} & {35.7} & {45.8} & {79.6} \\
\hline
\end{tabular}
\caption{Test-case considered, along with its dimensions.}
\label{fig:test_case4}
\end{figure}
In addition, with the aim of presenting the computational time-savings of the LATIN method, the results will be compared against a  classical step-by-step resolution, where Newmark scheme ($\gamma = 1/2$, $\beta=1/4$) is employed for time-integration and a Newton-Raphson scheme for handling the material nonlinearity. 
More precisely, the equilibrium residual is minimized using a quasi-Newton strategy where a preliminary Cholesky decomposition of the associated stiffness operator to invert is performed in order to compare the performances of the proposed developments with a classical but robust and powerful algorithm. To measure the error between the LATIN-PGD and Newmark/Newton-Raphson (NNR) solution the following error indicator is defined:
\begin{equation}\label{eq:error_L_NR}
\epsilon = 100 \sqrt{ \left( \frac{\norm{ \ctens{\sigma}_{\text{NNR}} - \ctens{\sigma}_{\text{LATIN}}}^2}{ \norm{\ctens{\sigma}_{\text{NNR}}}^2 } +
 \frac{\norm{ \ctens{\eps}_{\text{NNR}} - \ctens{\eps}_{\text{LATIN}}  }^2}{ \norm{\ctens{\eps}_{\text{NNR}}}^2 }
 \right) }  [\%]
\end{equation}
with $\norm{\ctens{\bullet}}^2 = \int_{\DS \times \DT} (\ctens{\bullet}) : (\ctens{\bullet}) \ d\DS dt$.

The variables $\ctens{\sigma}_{\text{NNR}}$, $\ctens{\sigma}_{\text{LATIN}}$, $\ctens{\eps}_{\text{NNR}}$ and $\ctens{\eps}_{\text{LATIN}}$ correspond to the stress and total deformation tensors obtained by the Newmark/Newton-Raphson and LATIN-PGD solver.

Regarding the temporal grid, the interval $\Cro{0;T}$ is split into $\NT$ coarse intervals. We parameterize the NNR algorithm to involve $2\NT+1$ time-steps adding an extra-node on each interval. The TDGM used in the LATIN-PGD approach involves Lagrange polynoms of degree three, yielding to $4\NT$ temporal DOFs.


The numerical test presented in this section involves isotropic damage for concrete material. 
The LATIN threshold considered for convergence is chosen in order to ensure good results of the LATIN-PGD solution and its value depend on the treated problem. 
The Table \ref{tab:error_threshold} on the other hand summarizes the threshold considered for the NNR method:
\begin{table}[H]
\begin{center}
\begin{tabular}{|c|}
\hline
NNR error (norm of the equilibrium)\\
\hline \hline
$10^{-4}$ \\
\hline
\end{tabular}
\caption{Error threshold considered for the NNR method.}\label{tab:error_threshold}
\end{center}
\end{table}
\vspace*{-\baselineskip}
\textbf{Important Remark}: In contrast to the LATIN error, which measure the distance between the local and global solutions manifolds, the error threshold of the NNR method is determined by computing a norm over the equilibrium equation. Both solvers use different error definitions, which makes it difficult to compare them under equal conditions. 
In this context, the error thresholds considered for both solvers were chosen based on empirical results only. Further research should be conducted to relate both errors in order to compare both solvers under equivalent conditions.
The following CPU times must then be considered with care.

In the following sub-sections, two displacement inputs are considered, a mono-periodic and a seismic like excitation, to load the beam of Figure \ref{fig:test_case4}.

\subsection{Simple mono-periodic excitation}

\begin{figure}[H] 
\centering
\includegraphics[width=0.45\textwidth]{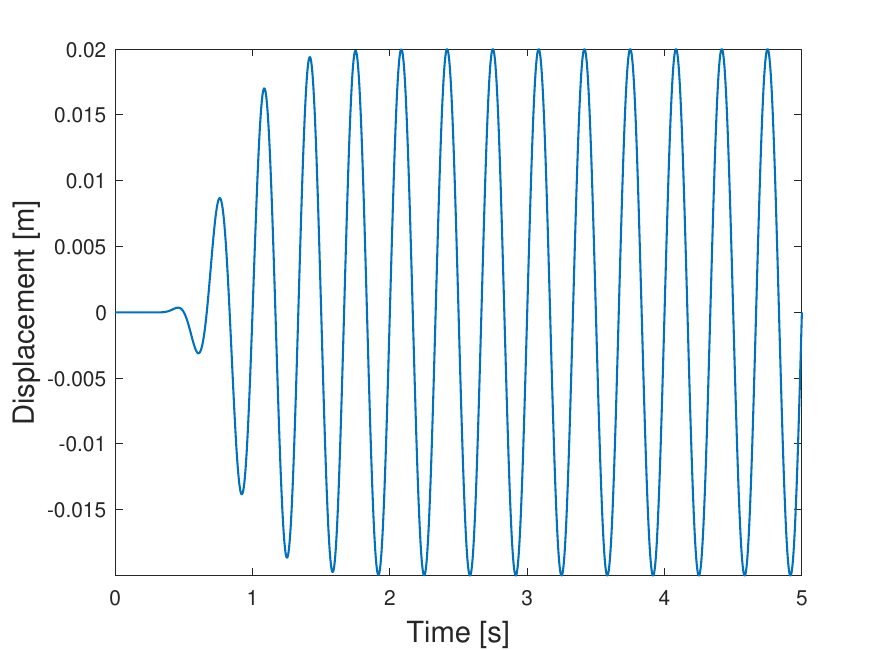}\caption{Imposed displacement.}
\label{fig:imp_displ_beton_4}
\end{figure}

\begin{figure}[H] 
\centering
\begin{subfigure}{0.48\textwidth}
\includegraphics[width=\textwidth]{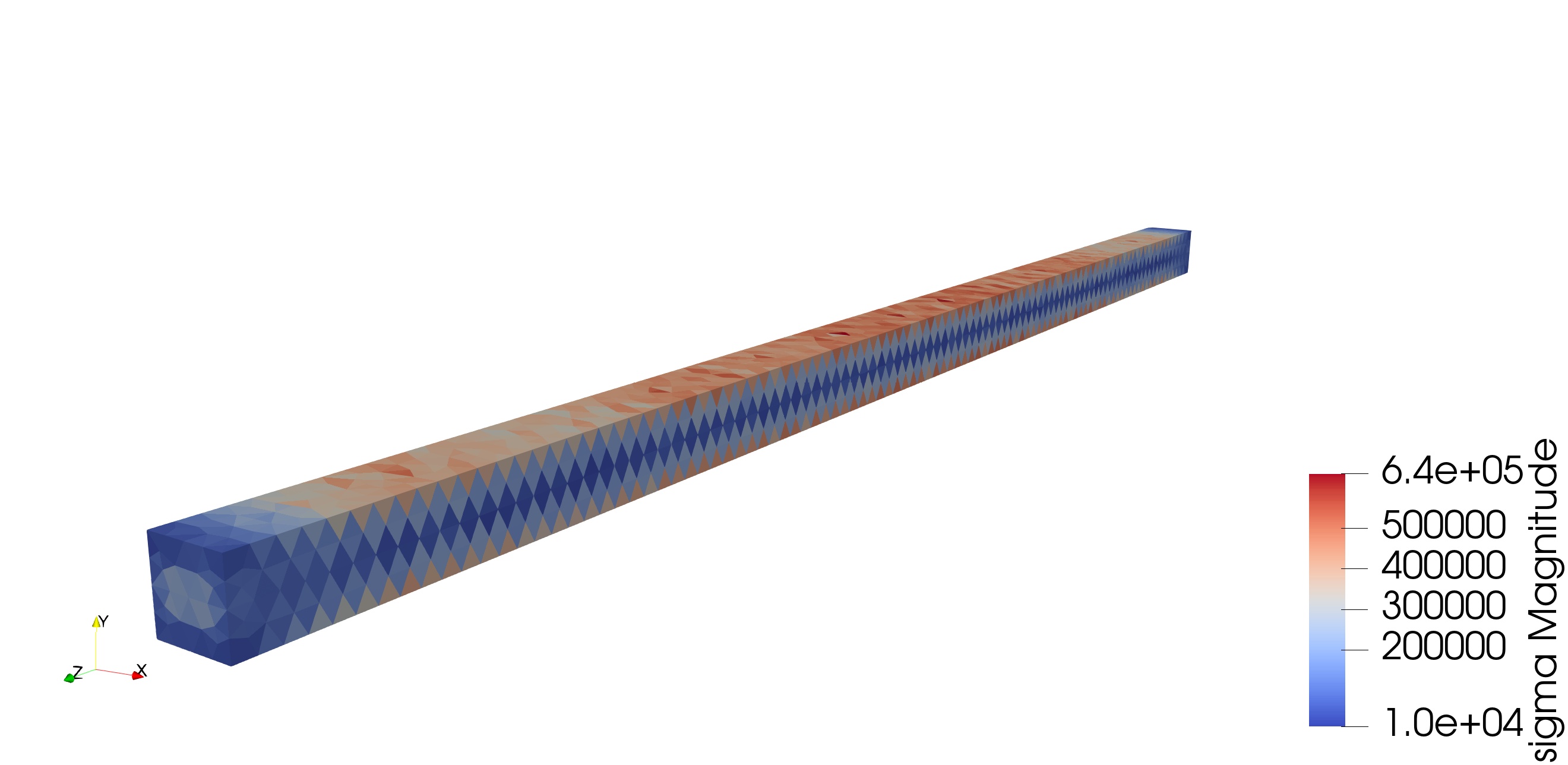}
\caption{Stress magnitude field obtained by the LATIN-PGD.}
\label{fig:beam_stress_LATIN}
\end{subfigure}
\begin{subfigure}{0.48\textwidth}
\includegraphics[width=\textwidth]{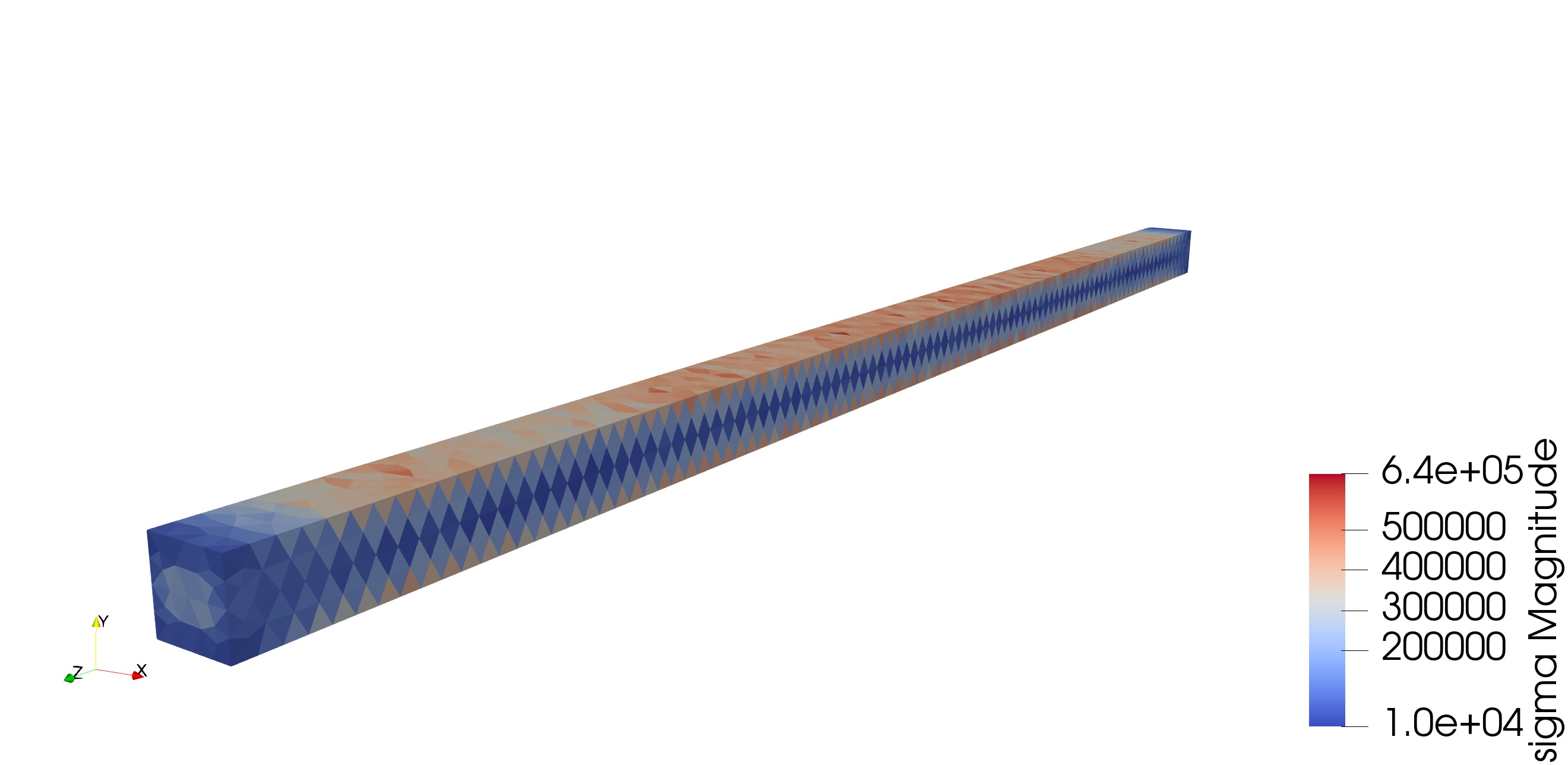}
\caption{Stress magnitude field obtained by the NR.}
\label{fig:beam_stress_NR}
\end{subfigure}
\caption{Magnitude of the stress tensor obtained with the LATIN-PGD (a) and NNR (b) at the end of the temporal domain ($t=T$).}
\label{fig:beam_stress}
\end{figure}

The prescribed vertical displacement $u_{z}^{D}(t)$ considered for this test is shown in Figure \ref{fig:imp_displ_beton_4} and consists of a sinusoidal excitation of frequency $3Hz$.
Under this condition the magnitude of the stress tensor field on the beam at the end of the time interval ($t=T$) given for the LATIN-PGD and the NNR method is presented in Figure \ref{fig:beam_stress}.

\begin{figure}[!ht] 
\centering
\begin{subfigure}{0.49\textwidth}
\includegraphics[width=\textwidth]{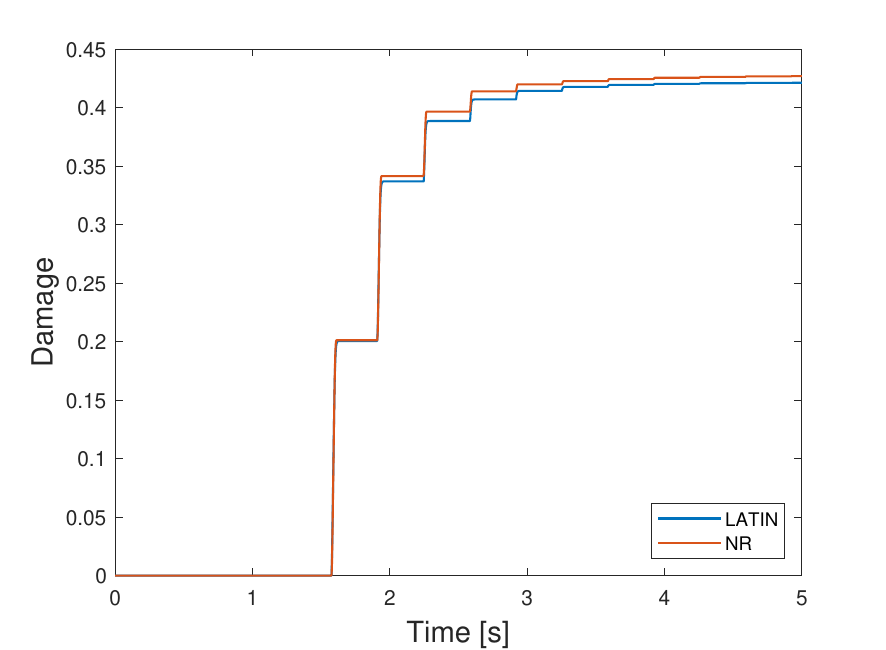}
\caption{}
\label{fig:dam_LATIN_vs_NR}
\end{subfigure}
\begin{subfigure}{0.49\textwidth}
\includegraphics[width=\textwidth]{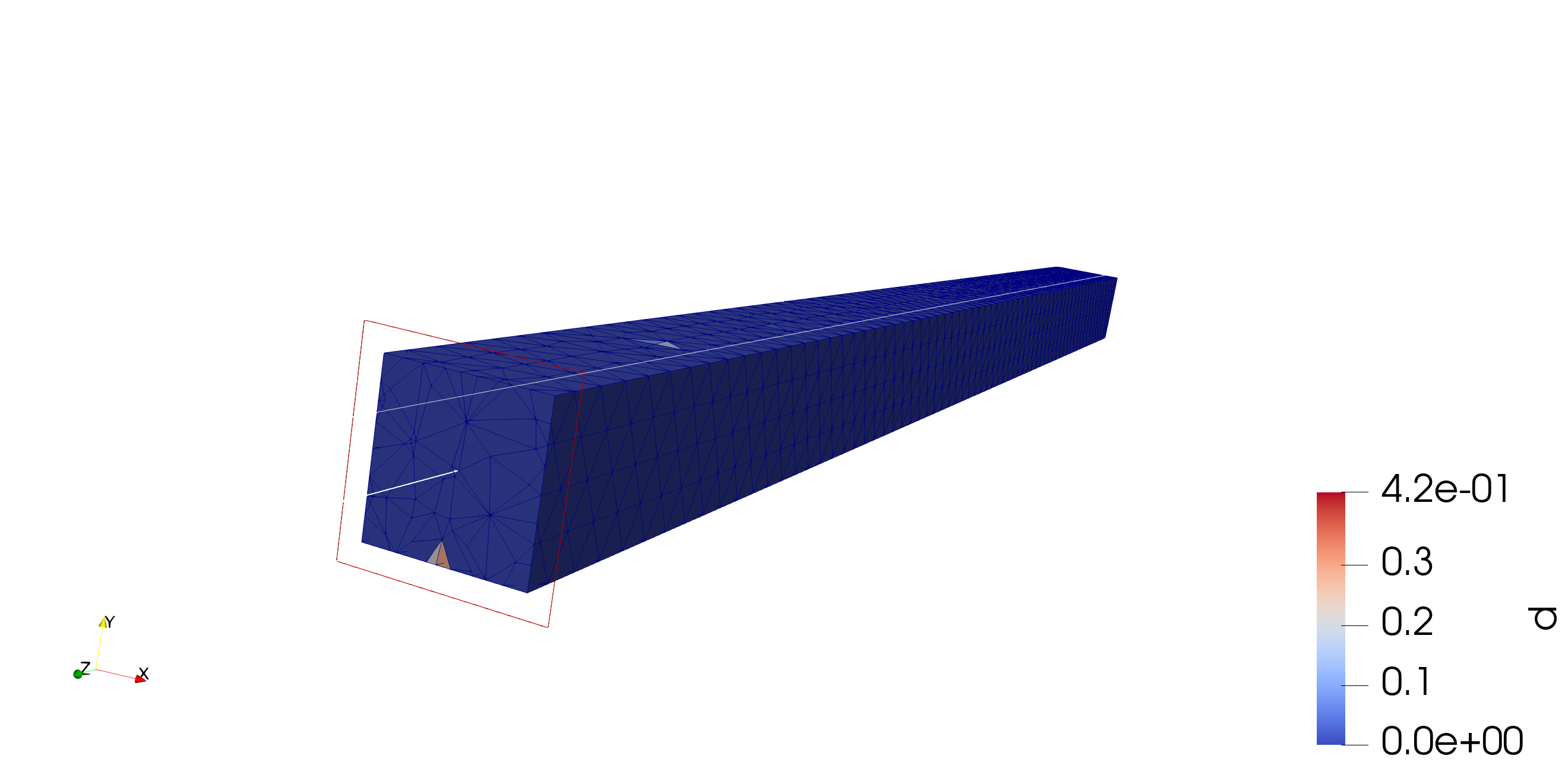}
\caption{}
\label{fig:damage_map}
\end{subfigure}
\caption{(a) Damage comparison between LATIN-PGD and NNR method; (b) Final damage map obtained by the LATIN-PGD method.}
\label{fig:iso_plots}
\end{figure}

Both computations are in very good accordance as can be seen on Figure \ref{fig:dam_LATIN_vs_NR} where the evolution of the damage variable over the most requested integration point in space is depicted. 
The values of maximum damage achieved in the same Gauss point are gathered in table \ref{tab:damage}.
\begin{table}[H]
\begin{center}
\begin{tabular}{|c|c|}
\hline
Solver & Damage \\
\hline \hline
NNR & 0.4212 \\
\hline
LATIN-PGD & 0.4269 \\
\hline
\end{tabular}
\caption{Comparison of maximal damage (obtained at same Gauss point).}\label{tab:damage}
\end{center}
\end{table}
\vspace*{-\baselineskip}

%
%

\begin{figure}[H] 
\centering
\includegraphics[width=0.5\textwidth]{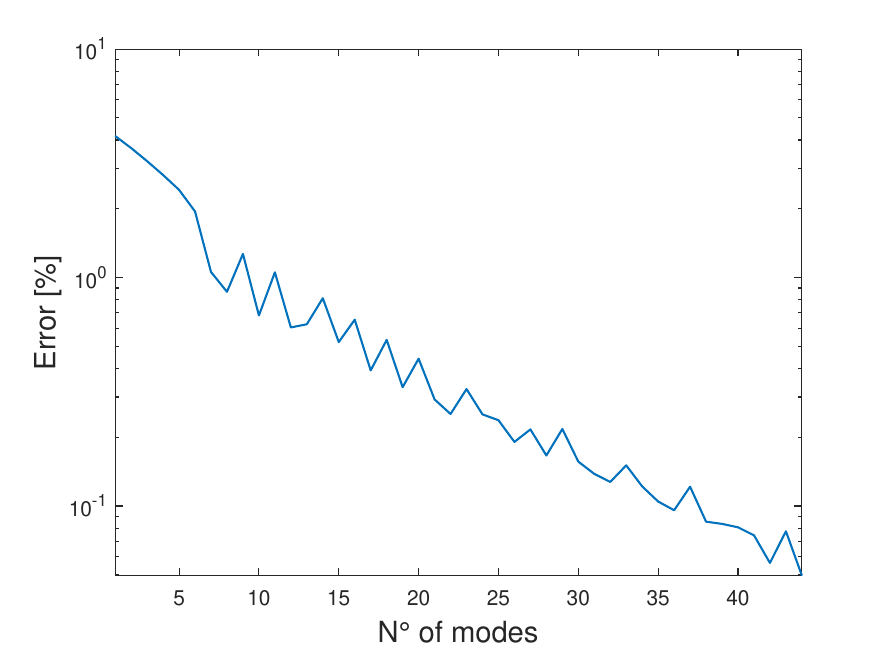}\caption{LATIN error (Log scale) vs PGD modes.}
\label{fig:LATIN_error_beton}
\end{figure}

The graph of the LATIN error vs the number of modes computed for this first example is shown in Figure \ref{fig:LATIN_error_beton}.
For this particular problem, a total of $45$ space-time PGD functions is required to reach a relative error $\LATINerror = 0.05 [\%]$. Finally, by using the error indicator $\epsilon$ defined in \eqref{eq:error_L_NR} between the LATIN-PGD and NNR solvers we obtained a relative difference of $0.8$ [\%].

\subsection{Multi-sine excitation}

\begin{figure}[H] 
\centering
\includegraphics[width=0.45\textwidth]{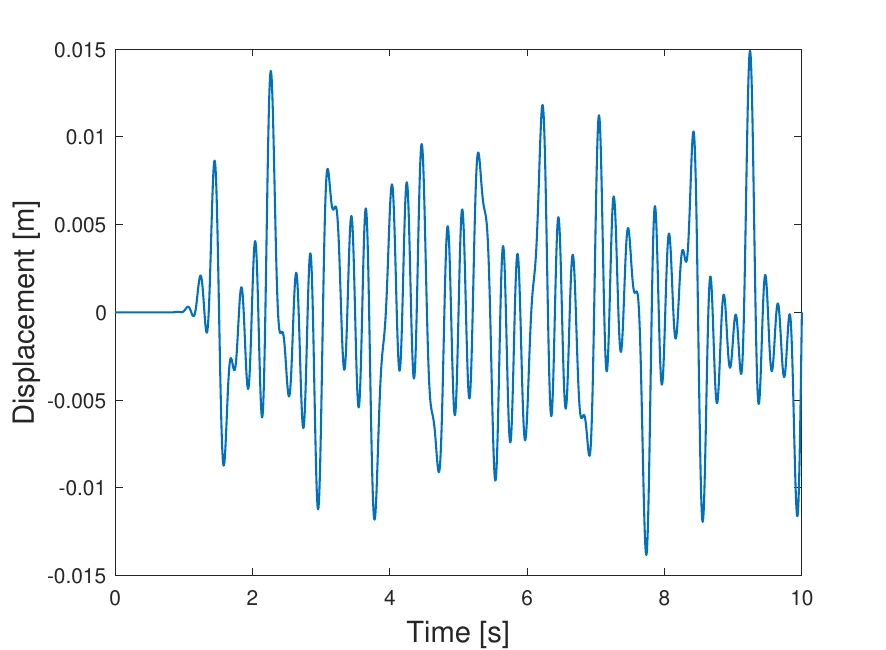}\caption{Imposed displacement.}
\label{fig:imp_displ_beton_rich}
\end{figure}

For this numerical test we consider a richer input $u_{z}^{D}(t)$ made of four sinusoidal components of frequencies 1, 2.3, 3.6 and \SI{5}{Hz} as depicted Figure \ref{fig:imp_displ_beton_rich}.
The elastic constitutive parameters of the beam are still those referred to Fig. \ref{fig:test_case4}.

Once again, in this second example, both LATIN-PGD and classical NNR computations are very close with a relative error level on the space-time domain $\epsilon = 1.5 [\%]$ as can be appreciated Figure \ref{fig:dam_LATIN_vs_NR_rich}. Note that compared to the first example, the LATIN convergence threshold has been moved to $\LATINerror=0.4 [\%]$ for appreciating the quality of the obtained solution when convergence threshold is relaxed and 28 PGD modes are necessary to achieve this solution.
\begin{figure}[!ht] 
\centering
\begin{subfigure}[t]{0.46\textwidth}
\includegraphics[width=\textwidth]{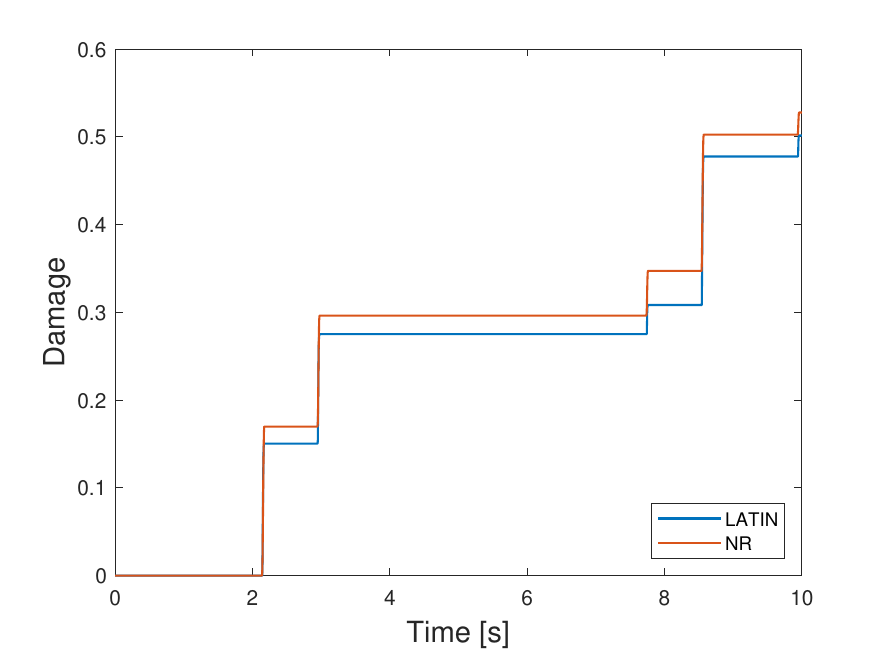}
\caption{Comparison of the damage obtained with the LATIN-PGD and the NNR methods on the most damaged Gauss point for the multi-sine case.}
\label{fig:dam_LATIN_vs_NR_rich}
\end{subfigure}
\hspace{12pt}
\begin{subfigure}[t]{0.46\textwidth}
\includegraphics[width=\textwidth]{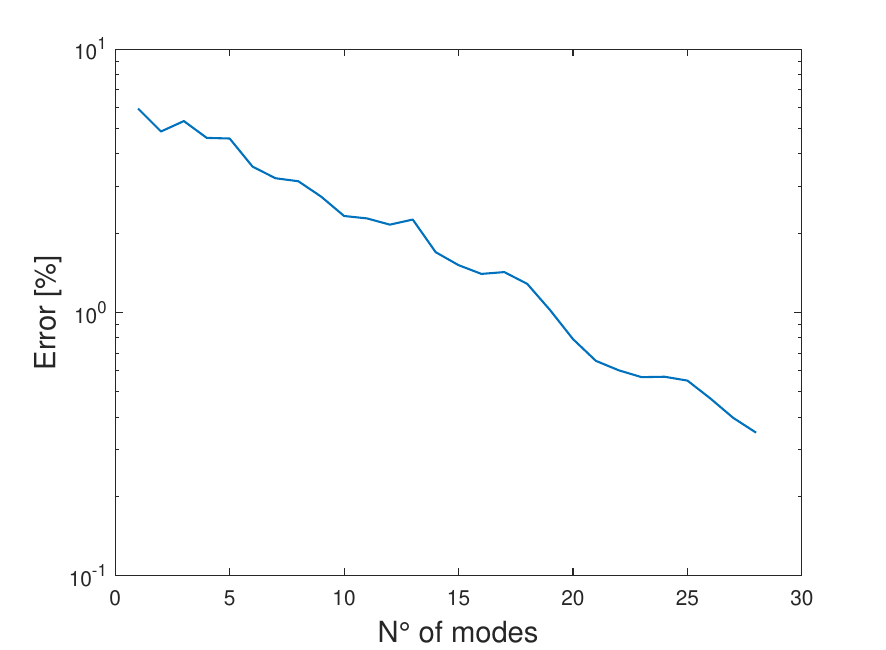}
\caption{LATIN error  $\LATINerror$ (Log scale) vs number of required PGD modes.}
\label{fig:dam_LATIN_vs_NR_rich}
\end{subfigure}
\caption{Numerical results for the multi-sine case.}
\label{fig:numres_rich}
\end{figure}

The total DOFs related to the reference problem for this example are resumed in Table \ref{tab:beton_test_dofs_rich}.
\begin{table}[H]
\begin{center}
\begin{tabular}{|c|c|c|c|}
\hline
Solver & Space DOFs & Time DOFs & Total DOFs \\
\hline \hline
NNR & 6471 & 2001 & 12948471 \\
\hline
LATIN-PGD & 6471 & 8000 & 51768000  \\
\hline
\end{tabular}
\caption{DOFs of the reference problem.}\label{tab:beton_test_dofs_rich}
\end{center}
\end{table}

Regarding CPU times, convergence is achieved in a few minutes for both solvers on a simple laptop without noting a broad superiority of the LATIN-PGD approach on this example.

%

\subsection{Conclusions on the numerical results}\label{sec:con_numeric}

This first implementation of the LATIN-PGD method involving a simple 3D concrete beam in flexion shows encouraging results.
A slight general CPU acceleration-rate has been observed on the two examples in favor of the LATIN-PGD strategy proposed in this paper.
However, stop criteria and convergence thresholds have been chosen using engineering judgment such as guaranteeing solution of overall equivalent `quality' for both NNR and LATIN solutions, and this, with more or less success as the plot of Fig. \ref{fig:dam_LATIN_vs_NR_rich} can testify.
An in-depth comparison of CPU times involving rigorous convergence threshold calibration for both NNR and LATIN-PGD approaches still need to be done and will be the object of future work.
Indeed, comparing simulation times is not an easy task; classical step-by-step NNR method (space-time local corrections) and LATIN-PGD method (space-time global corrections) compute their corrections in different manner.
One needs to make sure that the solutions have the same overall quality before performing a fair CPU-time comparison.
Results of such a benchmark could also depend on the studied quantity of interest (local failure index or more global damage quantity).
Nevertheless, even if an in-depth calibration of convergence thresholds should be done, the results presented in sections \ref{sec:Num_Results} leave good hopes for the improvement of convergence times and this for single computation, and even more when considering parametric studies (reuse of basis, clever initialization of the solution, etc.).

Note that gaining CPU-time on a single computation is not really the purpose of the work initiated in this paper.
Indeed, when considering seismic risk assessment, a large number of input have to be considered for computing seismic fragility curves and the LATIN approach, with its PGD-basis that can be reused, compressed and enriched on the fly, when computations progress brings an interesting framework for cleverly chaining the seismic calculations.
For instance, the two presented examples involved respectively 45 and 28 PGD modes that can be efficiently compressed to 9 and 10 modes using a truncated SVD.
Such truncated basis can be the starting point of a new calculation involving different load level or constitutive parameters.

Finally, the use of the Time-Discontinuous Galerkin method for the calculation of the PGD time functions allows for the considered numerical tests an interesting reduction of the size of the operators to be inverted to calculate these functions. 
Indeed, for the particular case of concrete, due to the choice of the constant operator $\HookeC$ as descent search direction, only one inversion of a single matrix of size $4 \times 4$ has been needed for the determination of the time functions in the enrichment step. These matrices are computed once and then reused over the entire temporal discretized domain, which allows to determine all temporal degrees of freedom by operating a single matrix of reduced size. This great advantage for the determination of temporal PGD functions makes the incremental TDGM solver a powerful tool for the LATIN-PGD method, especially when a large number of temporal DOFs must be determined.

\section{Conclusions}\label{sec:concl_persp}

This paper presents a first implementation of the LATIN-PGD methodology for solving a nonlinear problem in low-frequency dynamics. 

Our underlying objective is to lay the corner-stone for handling the uncertainties and dealing with complex parametric studies encountered for seismic risk assessment: fragility curves in particular impose the computation of a structure with respect to a family of likely seismic inputs and the model-reduction framework brought by the LATIN-PGD approach naturally favors numerical efficiency when solving such kind of quasi-identical problems.
The methodology is presented for the case concrete material modeled using isotropic damage description, but other types of materials could be considered without major difficulty. 
The numerical examples, highlight the potentialities of the proposed methodology and brought out the computational gain that was enabled by our approach when estimating the mechanical response of the structure in comparison with a classical incremental solver. 
Such reduction is directly proportional to the number of degrees of freedom considered for both spatial and temporal domain. The last is achieved because the LATIN method is a non-incremental solver that enables the use of the model-order reduction PGD to express the nonlinear solution of the problem as a low-rank approximation. 
The PGD allows to capture the redundancies of the nonlinear solution in both space and time, allowing a considerable reduction of computation time.

In addition, an incremental strategy based on the Time-Discontinuous Galerkin Method (TDGM) was introduced, which allows to efficiently compute the temporal PGD functions, since it allows to reduce the size of the matrices related to the temporal discretized problems needed to be inverted for the enrichment step of the LATIN-PGD method. 
This technique is very useful and imposes an advantage over the classically used methods, which correspond to a continuous formulation in time using the Time Continuous Galerkin Method (TCGM), especially when the time domain is large. 
In these situations, the continuous formulation requires the assembly and inversion of large matrices for the temporal resolution, which decreases the efficiency of the LATIN-PGD method. 
The discontinuous TDGM formulation however only leads to assemble small element-by-element temporal operators, avoiding the construction and inversion of large assembled matrices, thus increasing the efficiency of the LATIN-PGD method.

\begin{acknowledgements}

The authors are grateful to the ``Commissariat à l'énergie atomique et aux énergies alternatives" (CEA) and the ``Institut de radioprotection et de sûreté nucléaire" (IRSN) for funding this research.
\end{acknowledgements}

%
\section*{Conflict of interest}
The authors declare that they have no conflict of interest.

\bibliographystyle{spmpsci}      
\bibliography{bibliography}   

\begin{appendices}

\include{AppendixA}

\end{appendices}

\end{document}

%% file: Colors.tex

\definecolor{myorange}{rgb}{0.95, .7, .45}
\definecolor{Myorange}{rgb}{0.95, .7, .45}

\definecolor{myred}{rgb}{0.95, .7, .7}
\definecolor{Myred}{rgb}{0.90, .5, .5}

\definecolor{Myyellow}{rgb}{0.95, 0.95, .45}
\definecolor{myyellow}{rgb}{1, 1, .7}

\definecolor{Mygreen}{rgb}{0.2, 0.6, 0.2}
\definecolor{mygreen}{rgb}{0.2, 0.6, 0.2}

\definecolor{Mygray}{rgb}{0.15, 0.15, 0.15}

\definecolor{myblue}{rgb}{0.8, 0.8, 1}
\definecolor{Myblue}{rgb}{0.8, 0.8, 1}
\definecolor{NavyBlue}{rgb}{0.29, 0.49, 0.73}             
\definecolor{RoyalBlue}{rgb}{0.29, 0.19, 0.43}

\definecolor{qqwwtt}{rgb}{0.,0.4,0.2}     
\definecolor{ccqqqq}{rgb}{0.8,0.,0.}      
\definecolor{qqttzz}{rgb}{0.,0.2,0.6}     

\definecolor{mycolor}{rgb}{0.,0.,0.}      


%% file: Commandes.tex
\newcommand{\R}{\mathbb{R}}

\newcommand{\Ctens}[1]{%
            \underline{\mathit{%
                {#1}
            }}
}




\newcommand{\SDtens}[1]{%
            \underline{\underline{\boldsymbol{\mathbf{%
                \mathit{#1}
            }}}}
}

\newcommand{\SDvect}[1]{%
            \underline{\boldsymbol{\mathbf{%
                \mathit{#1}
            }}}
}



\newcommand{\Acc}[1]{\left\{{#1}\right\}}
\newcommand{\ACC}[1]{\Big\{{#1}\Big\}}
\newcommand{\cro}[1]{[{#1}]}
\newcommand{\Cro}[1]{\left[{#1}\right]}
\newcommand{\CRO}[1]{\Big[{#1}\Big]}
\newcommand{\prth}[1]{({#1})}
\newcommand{\Prth}[1]{\left({#1}\right)}
\newcommand{\PRTH}[1]{\Big({#1}\Big)}

\newcommand{\abs}[1]{\left|#1\right|}

\newcommand{\norm}[1]{\left\|{#1}\right\|}

\newcommand{\eq}{ \ = \ }
\newcommand{\deq}{ \ \triangleq \ }

\newcommand{\pbic}{:}             


\newcommand{\x}{.}                




\newcommand{\VVsig}{\Ctens{\Ctens{\sigma}}}
\newcommand{\VVeps}{\Ctens{\Ctens{\varepsilon}}}


\newcommand{\DS}{{\Omega}}             
\newcommand{\DT}{{I}}                  
\newcommand{\DTd}{\breve{\DT}}





\newcommand{\IV}{{X}}
\newcommand{\IVY}{Y}
\newcommand{\IVd}{d}
\newcommand{\dIVd}{\dot{d}}
\newcommand{\IVZ}{Z}
\newcommand{\IVz}{z}

\newcommand{\MPAd}{A_d}
\newcommand{\MPYs}{Y_0}
\newcommand{\MPtauC}{\tau}
\newcommand{\MPaC}{a}

\newcommand{\HookeC}{\mathbb{E}}

\newcommand{\mbraP}[1]{\langle{#1}\rangle_{+}} 			
\newcommand{\MbraP}[1]{\left\langle{#1}\right\rangle_{+}}	

\newcommand{\fDth}{f} 		

\newcommand{\Fcc}{\mathcal{F}} 		

%

\newcommand{\Eu}{\mathscr{U}}        
\newcommand{\EuS}{\Eu^S}             
\newcommand{\Esig}{\mathscr{S}}

\newcommand{\Eeps}{\mathscr{E}} 	

\newcommand{\Esobolev}{\mathscr{H}_1}


\newcommand{\NgS}{N_g^S}
\newcommand{\NgT}{N_g^T}


\newcommand{\RdC}{\mathcal{R}}

\newcommand{\SOL}{\mathcal{S}}
\newcommand{\EGamma}{\mathbf{\Gamma}}
\newcommand{\EAd}{\mathbf{Ad}}
\newcommand{\Egamma}{\EGamma}

\newcommand{\SDGA}{\mathbb{A}}         
\newcommand{\SDAG}{\mathbb{G}}

\newcommand{\na}{(n)}
\newcommand{\nb}{(n+\rfrac{1}{2})}
\newcommand{\nc}{(n+1)}
\newcommand{\SOLinit}{\SOL^{(0)}}
\newcommand{\SOLa}{\SOL^{\na}}
\newcommand{\SOLb}{\hat{\SOL}^{\nb}}
\newcommand{\SOLc}{\SOL^{\nc}}

\newcommand{\SOLrelax}{\breve{\SOL}^{\nc}}
\newcommand{\VVsigh}{\hat{\VVsig}}
\newcommand{\VVsigCh}{\hat{\VVsig}_c}
\newcommand{\VVepsh}{\hat{\VVeps}}
\newcommand{\IVdh}{\hat{\IVd}}
\newcommand{\IVYh}{\hat{\IVY}}
\newcommand{\IVzh}{\hat{\IVz}}
\newcommand{\IVZh}{\hat{\IVZ}}




\newcommand{\pscT}[1]{\, \big\langle #1 \big\rangle_\DT \,}

\newcommand{\normST}[1]{\big|\big|\big|#1\big|\big|\big|}
\newcommand{\NormST}[1]{\left|\left|\left| #1 \right|\right|\right|}




\newcommand{\gS}{g_S}
\newcommand{\gT}{g_T}


\newcommand{\tableq}[2]{
\hspace{0.01\hsize}\makebox[0.2\hsize][l]{${#1}$} \hspace{-0.17\hsize}\makebox[0.88\hsize][c]{$\displaystyle #2$}
}



\newcommand{\Par}{\medskip\par}

\definecolor{myorange}{rgb}{0.95, .7, .45}
\definecolor{Myorange}{rgb}{0.95, .7, .45}

\definecolor{myred}{rgb}{0.95, .7, .7}
\definecolor{Myred}{rgb}{0.90, .5, .5}

\definecolor{Myyellow}{rgb}{0.95, 0.95, .45}
\definecolor{myyellow}{rgb}{1, 1, .7}

\definecolor{Mygreen}{rgb}{0.2, 0.6, 0.2}
\definecolor{mygreen}{rgb}{0.2, 0.6, 0.2}

\definecolor{Mygray}{rgb}{0.15, 0.15, 0.15}

\definecolor{myblue}{rgb}{0.8, 0.8, 1}
\definecolor{Myblue}{rgb}{0.8, 0.8, 1}
\definecolor{NavyBlue}{rgb}{0.29, 0.49, 0.73}             
\definecolor{RoyalBlue}{rgb}{0.29, 0.19, 0.43}

\definecolor{qqwwtt}{rgb}{0.,0.4,0.2}     
\definecolor{ccqqqq}{rgb}{0.8,0.,0.}      
\definecolor{qqttzz}{rgb}{0.,0.2,0.6}     

%% file: Notations.tex
\newcommand{\test}[1]{{#1}^{\ast}}




\newenvironment{parboxed}[1][black]
  {\def\FrameCommand{\fboxsep=\FrameSep\fcolorbox{#1}{white}}%
    \MakeFramed {\advance\hsize-\width \FrameRestore}}
  {\endMakeFramed}


\theoremstyle{definition}
\newtheorem{prob}{\textsc{Problem}}
\newenvironment{probleme}[1][htbp]{
		\begin{parboxed}
		    \begin{prob}
	}{
		    \end{prob}
		\end{parboxed}
	}





\newcommand{\rmDelta}{\mathrm{\Delta}} 

\newcommand{\eqtext}[1]{{\vspace{2mm} \noindent {#1}}}

\newcommand{\com}{,} 

\newcommand{\dotp}{\cdot} 

\newcommand{\p}{\hspace{0.5mm}} 


\newcommand{\SOLex}{\bar{\SOL}}

\newcommand{\med}{me}
\newcommand{\cra}{cr}

\newcommand{\rhoc}{\rho}



\newcommand{\vect}[1]{\underline{{#1}}} 

\newcommand{\ctens}[1]{\underline{\underline{{#1}}}} 














\newcommand{\eps}{\varepsilon}









\newcommand{\normi}[1]{\left\|{#1}\right\|}







\newcommand{\Ym}{E}

\newcommand{\Pc}{\nu}



\newcommand{\Ymc}{\Ym}
\newcommand{\Pcc}{\Pc}

\newcommand{\Y}{Y}
\newcommand{\Ad}{A_{d}}
\newcommand{\dd}{\tau_{c}}
\newcommand{\ddexp}{a}
\newcommand{\cc}{a_{c}}








































\newcommand{\sfT}{\psi}
\newcommand{\sfTv}{\vect{\sfT}}

\newcommand{\vx}{\vect{x}} 











\newcommand{\DTk}{\breve{\DT}_{k}} 


\newcommand{\Su}{\mathcal{U}} 
\newcommand{\SuS}{\Su^S}  
\newcommand{\SuT}{\Su^T}  





 %



             
            
\newcommand{\NT}{N_T}              

\newcommand{\NS}{N_S}

\newcommand{\Npgt}{n_{t,g}}




\newcommand{\init}{{0}}


\newcommand{\CREc}{J}


















\newcommand{\ie}{(0)}

\newcommand{\uo}{\vect{u}^{\ie}}

\newcommand{\sigmao}{\ctens{\sigma}^{\ie}}
\newcommand{\epso}{\ctens{\varepsilon}^{\ie}}










\newcommand{\pgderror}{\zeta_{PGD}}


\newcommand{\LATINerror}{\xi}



\newcommand{\SDa}{\mathit{G}}
\newcommand{\SDd}{\mathit{A}}

\newcommand{\SDaC}{\SDa}
\newcommand{\SDdC}{\SDd}









\newcommand{\LM}{\SDtens{\mathcal{L}}}

\newcommand{\RM}{\SDtens{\mathcal{R}}}

\SetKwInOut{Parameter}{Parameters}

\newcommand{\utest}{\test{\vect{u}}}

\newcommand{\utestdtdt}{\test{\ddot{\vect{u}}}}

\newcommand{\PGDm}{m} 


\newcommand{\pgdv}{\test{\bar{\vect{u}}}} 
\newcommand{\pgdu}{\bar{\vect{u}}}

\newcommand{\pgdsig}{\bar{\ctens{\sigma}}}

\newcommand{\pgdeps}{\bar{\ctens{\varepsilon}}}

\newcommand{\Lres}{{\ctens{\rmDelta}}}


\newcommand{\pgdt}{\lambda}




\newcommand{\TU}{\rmDelta \vect{\lambda} }



\newcommand{\Mt}{\SDtens{Q}}
\newcommand{\Ft}{\SDvect{f}}























































\newcommand{\ntg}{\Npgt}
\newcommand{\nsg}{n_{s,g}}























\renewcommand{\SOLex}{\SOL}

%% file: AppendixA.tex
\label{sec:appendix_enrich}